\documentclass[12pt,thmsb,titlepage]{article}
\usepackage{amsmath}
\usepackage{amssymb}
\usepackage{graphicx}
\usepackage{epsf}
\usepackage{epstopdf}

\newtheorem{theorem}{Theorem}

\newtheorem{lemma}{Lemma}
\newtheorem{definition}{Definition}
\newtheorem{corollary}{Corollary}
\newtheorem{ass}{Assumption}

\newcommand{\captionfonts}{\footnotesize}
\makeatletter  
\long\def\@makecaption#1#2{%
  \vskip\abovecaptionskip
  \sbox\@tempboxa{{\captionfonts #1: #2}}%
  \ifdim \wd\@tempboxa >\hsize
    {\captionfonts #1: #2\par}
  \else
    \hbox to\hsize{\hfil\box\@tempboxa\hfil}%
  \fi
  \vskip\belowcaptionskip}
\makeatother   

\begin{document}

\centerline{\large{Optimal Decision Rules for Weak GMM}}

\centerline{By Isaiah Andrews\footnote{ Harvard Department of Economics, Littauer Center M18, Cambridge, MA 02138. Email iandrews@fas.harvard.edu.  Support from the National Science Foundation under grant number 1654234, and from the Sloan Research Fellowship is gratefully acknowledged.}
 and Anna Mikusheva\footnote{
Department of Economics, M.I.T., 50 Memorial Drive, E52-526, Cambridge, MA, 02142. Email: amikushe@mit.edu.
      Financial support from the Castle-Krob
      Career Development Chair and the Sloan Research Fellowship is gratefully acknowledged.  We thank Jiafeng Chen, David Hughes, and Ashesh Rambachan for research assistance, and the editor, two anonymous referees, and numerous seminar participants for helpful comments.}}

\centerline{Abstract}

 {\small{
 This paper studies optimal decision rules, including estimators and tests, for weakly identified GMM models.  We derive the limit experiment for weakly identified GMM, and propose a theoretically-motivated class of priors which give rise to quasi-Bayes decision rules as a limiting case.  Together with results in the previous literature, this establishes desirable properties for the quasi-Bayes approach regardless of model identification status, and we recommend quasi-Bayes for settings where identification is a concern. We further propose weighted average power-optimal identification-robust frequentist tests and confidence sets, and prove a Bernstein-von Mises-type result for the  quasi-Bayes posterior under weak identification.
 }
 }

\begin{center}
First draft: July 2020.

This draft: July 2021.
\end{center}

\section{Introduction}

Weak identification arises in a wide range of empirical settings.  Weakly identified nonlinear models have objective functions which are near-flat in certain directions, or have multiple near-optima.  Standard asymptotic approximations break down when identification is weak, resulting in biased and non-normal estimates, as well as invalid standard errors and confidence sets.  Further, existing optimality results do not apply in weakly identified settings, making it unclear how researchers should best extract the information present in the data.

To provide guidance for such cases, this paper develops a theory of optimality for nonlinear GMM models with weak identification.\footnote{Kaji (2020) studies weakly identified parameters in semiparametric models, and introduces a notion of weak efficiency for estimators. Weak efficiency is necessary, but not in general sufficient, for decision-theoretic optimality (e.g. admissibility) in many contexts.}  Even when identification is weak the sample average GMM moment function is approximately normally distributed, allowing us to derive a novel Gaussian process limit experiment which upper-bounds attainable large-sample performance.
This limit experiment is infinite-dimensional, reflecting the semiparametric nature of the GMM model, and there typically exists no uniformly best procedure.  As a consequence, asymptotically optimal approaches for weakly identified GMM necessarily trade off performance across different parts of the parameter space.

We discipline these tradeoffs by considering Bayes decision rules, and propose a theoretically-motivated class of computationally tractable priors for the limit experiment.  This class yields the quasi-Bayes approach studied by Kim (2002) and Chernozhukov and Hong (2003) as a diffuse-prior limit.  Quasi-Bayes treats a transformation of the continuously updating GMM objective function as a likelihood, which is combined with a prior to produce Bayes decision rules.  We further prove a Bernstein-von Mises-type result establishing the asymptotic properties of  quasi-Bayes under weak identification.

For inference, one may report quasi-Bayes credible regions.  However, these do not in general have correct frequentist coverage under weak identification.  To address this issue, we derive weighted average power-maximizing frequentist tests with respect to our recommended priors, and construct confidence sets by collecting the parameter values not rejected by these tests.  These confidence sets take a form similar to quasi-Bayes highest posterior density regions, but ensure correct coverage under weak identification.

Finally, we illustrate our analysis with simulations and empirical results based on quantile IV applied to data from Graddy (1995) on the demand for fish.  

The bottom line of this analysis is that quasi-Bayes decision rules have desirable optimality properties whether identification is weak or strong.  Unlike for conventional GMM estimators, to use the quasi-Bayes approach researchers must specify a prior.  Chernozhukov and Hong (2003) show, however, that quasi-Bayes procedures are equivalent to efficient GMM in large samples under strong identification, and so are insensitive to the prior in this case.  Under weak identification, by contrast, the prior plays an important role, since the data provide only limited information.  Even under weak identification, however, our robust confidence sets guarantee correct coverage regardless of the choice of prior.  These results complement earlier findings from Chen, Christensen, and Tamer (2018) who show, among many other findings, that quasi-Bayes highest posterior density sets based on flat priors have correct coverage for the identified set in strongly but partially identified models, establishing a further desirable property for quasi-Bayes approaches in settings with non-standard identification.\footnote{Chen, Christensen, and Tamer (2018) also show that, in the partially, strongly identified setting, quasi-Bayes posteriors based on informative priors can be used to form critical values for confidence sets for the full parameter vector and subvectors, and propose simple confidence sets for scalar subvectors.}
Based on these results, we recommend quasi-Bayes for settings where identification is a concern.

The next section introduces our setting and derives the limit experiment.  Section \ref{section- prior} motivates and derives our recommended class of priors and shows that quasi-Bayes corresponds to their diffuse limit.  Section \ref{sec: quasi-posterior rules} discusses quasi-Bayes decision rules and constructs optimal frequentist tests.  Section \ref{sect - implementation} discusses feasible quasi-Bayes decision rules and characterizes their large-sample behavior.  Finally, Section \ref{section: empirical} provides empirical results for quantile IV applied to data from Graddy (1995).

\section{Limit Experiment for Weakly Identified GMM}\label{sec: limit experiment}
\subsection{Weak Identification in Nonlinear GMM}\label{sec: weak ID}
Suppose we observe a sample of independent and identically distributed observations $\{X_i, i=1,...,n\}$ with support $\mathcal{X}$.
We are interested in a structural parameter $\theta^*\in\Theta$, which is assumed to satisfy the moment equality
$
\mathbb{E}_{P}\left[\phi(X,\theta^*)\right]=0\in\mathbb{R}^k,
$
for $P$ the unknown distribution of the data and $\phi(\cdot,\cdot)$ a known function of the data and parameters.\footnote{Our results in this section allow infinite-dimensional structural parameters, but in the remainder of the paper we assume that $\theta$ is finite-dimensional.}  We assume $\theta^*$ corresponds to a quantity of economic interest which is well-defined whether or not it can be recovered from the data, and want to choose an action $a\in\mathcal{A}$ to minimize a loss $L(a,\theta^*)\ge0$ that depends only on $a$ and $\theta^*$.

\paragraph{Example: Quantile IV} Suppose the observed data $X=(Y,W,Z)$ consist of an  outcome $Y$, a $(p-1)$-dimensional vector of endogenous regressors $W$, and a $k$-dimensional vector of instruments $Z$.
For $\theta=(\alpha,\beta)\in\Theta$, consider the moment condition
\begin{align}\label{eq: quantile IV}
\phi\left(X,\theta\right)=\left(\mathbb{I}\{Y-\alpha-W'\beta\leq 0\}-\tau\right)Z,
\end{align}
introduced by  Chernozhukov and Hansen (2005) for inference on the $\tau$-th conditional quantile. For estimation of $\theta$ we could take $\mathcal{A}=\Theta$ and consider squared error loss, $L(a,\theta)=\|a-\theta\|^2.$ $\square$

If the data are informative, standard asymptotic arguments imply that the GMM estimator for $\theta^*$ is approximately normally distributed in large samples, with standard errors that can be consistently estimated.  Under regularity conditions, one can further show that the optimally-weighted GMM estimator is asymptotically efficient, in the sense of minimizing expected squared error loss in large samples, and that efficient decision rules for many other problems may be constructed based on this estimator.   By contrast, when the data are less informative and identification is weak, these approximations break down.  Consequently, GMM estimators may be far from normally distributed, and it is less clear how to efficiently estimate $\theta^*$ or solve other decision problems.

\paragraph{Example: Quantile IV, continued}
We illustrate the problem of weak identification with an application
of quantile IV to data from Graddy (1995) on the demand for fish at
the Fulton fish market. Following Chernozhukov et al. (2009), who
discuss finite-sample frequentist inference in this setting, we consider
the quantile IV moment conditions  stated in equation (\ref{eq: quantile IV}) with $Y$ the log quantity of fish
purchased, $W$ the log price, and $Z$ a vector of instruments consisting
of a constant, a dummy for whether the weather offshore was mixed
(with wave height above 3.8 feet and windspeed over 13 knots), and
a dummy for whether the weather offshore was stormy (with wave height
above 4.5 feet and windspeed over 18 knots).
We focus on results for the 75th percentile, $\tau=0.75$.
For further details on
the data and setting, see Graddy (1995) and Chernozhukov et al. (2009).

\begin{figure}
\includegraphics[scale=0.5]{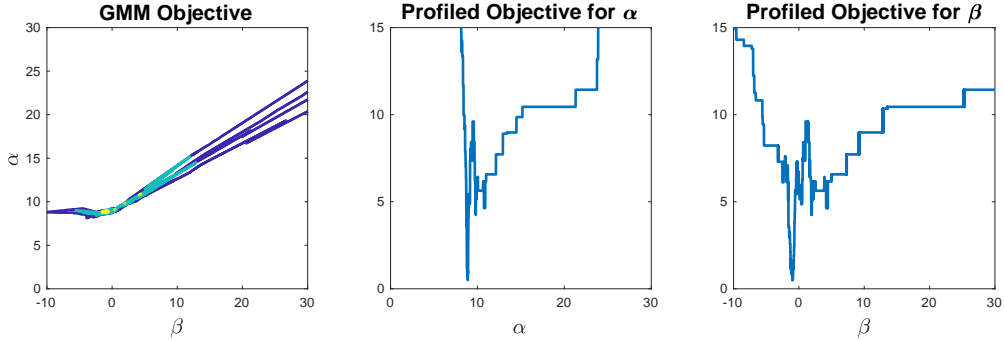}
\caption{GMM objective function $Q_n(\theta)$ for quantile IV, $\tau=0.75,$ based on Graddy (1995) data.\label{fig: GMM objective plot}}
\end{figure}

The first panel of Figure \ref{fig: GMM objective plot} plots contours of the continuously updating GMM objective function, $Q_n(\theta)=g_{n}\left(\theta\right)'\widehat{\Sigma}_n^{-1} \left(\theta,\theta\right)g_{n}\left(\theta\right)$, for $g_n(\cdot)=\frac{1}{\sqrt{n}}\sum_{i=1}^n\phi(X_i,\cdot)$ and $\widehat\Sigma_n(\theta,\theta)$ the sample variance of $\phi(X_i,\theta)$.  The second and third panels plot the profiled objective functions for $\alpha$ and $\beta$, respectively, where the profiled objective for $\alpha$ is $\min_\beta Q_n(\alpha,\beta)$.
The GMM objective is low over a substantial part of the parameter space suggesting, as previously noted by Chernozhukov et al. (2009), that weak identification may be an issue here.

The details of Figure \ref{fig: GMM objective plot} suggest particular trouble for conventional asymptotic approximations.  Standard arguments for the large-sample normality of GMM estimators rely on quadratic approximations to the objective function, but Figure \ref{fig: GMM objective plot} shows that the sample objective function is far from quadratic in these data.
To explore the implications for GMM estimation, we calibrate simulations to the Graddy (1995) data.\footnote{Details of the calibration may be found in  Appendix D.}  The first column of Figure \ref{fig: q75 approximation} shows the distribution of the GMM estimators for $\alpha$ and $\beta$ in these calibrations, and highlights that these distributions are clearly non-normal, with substantial right-skewness.  Hence, we see that weak identification undermines the validity of conventional asymptotic approximations in this setting, raising the question of how to efficiently estimate the structural parameters. $\square$

To develop asymptotic results for weakly identified GMM, we need a framework for modeling weak identification.
Intuitively, $\theta^*$ is weakly identified when the mean of the moment function $\mathbb{E}_{P}[\phi(X,\theta)]$ is close to zero, relative to sampling uncertainty, over a non-trivial part of the parameter space $\Theta$.  To obtain asymptotic approximations that reflect this situation, we adopt a nonparametric version of weak identification asymptotics and model the data generating process as local to identification failure.  Specifically, we assume the sample of size $n$ follows distribution $P=P_{n,f},$ where the sequence $P_{n,f}$ converges to some (unknown) limit $P_0$,
\begin{equation}
\int\left[\sqrt{n}(dP_{n,f}^{1/2}-dP_0^{1/2})-\frac{1}{2}fdP_0^{1/2}\right]^{2}\to0 
\label{eq: Path restriction}
\end{equation}
as $n\to\infty$. Restriction (\ref{eq: Path restriction}) is a differentiability in quadratic mean condition, and is standard in the literature on semi-parameteric efficiency (see Chapter 25 in van der Vaart, 1998).\footnote{Prior work by Kaji (2020) also analyzes weak identification using paths satisfying (\ref{eq: Path restriction}).}
We focus on the case where the structural parameter $\theta^*$ is set-identified under $P_0$, in the sense that the identified set
 \[
\Theta_{0}=\left\{ \theta\in\Theta:\mathbb{E}_{P_0}[\phi(X,\theta)]=0\right\}
\]
contains at least two distinct elements, and assume that $\Theta_0$ is compact, with $\theta^*\in\Theta_0$.\footnote{
The more general assumption that $\theta^*$ is local to $\Theta_0$ yields a limit experiment similar to that derived below, at the cost of heavier notation.  Hence, we focus on the case with $\theta^*\in\Theta_0$.}

 A measurable function $f$ in (\ref{eq: Path restriction}) is called a score, and necessarily satisfies $\mathbb{E}_{P_0}\left[f(X)\right]=0$ and $\mathbb{E}_{P_0}[f^2(X)]<\infty$ (see  Van der Vaart and Wellner, 1996, Lemma 3.10.10). Denote the space of score functions by $T(P_0)$, and note that this is a linear subspace of  $L_{2}(P_0)$, the space of functions from $\mathcal{X}$ to $\mathbb{R}$ that are square integrable with respect to $P_0$.
Intuitively, each $f\in T(P_0)$ specifies a different ``direction'' from which the sequence of data generating processes $P_{n,f}$ may approach $P_0$.  The space of such directions is typically infinite-dimensional (in particular, whenever $X$ has a continuously distributed element), and describes  the many ways in which the data generating $P_{n,f}$ may depart from $P_0$ in terms of means and covariances, but also in terms of higher moments and other features.  Condition (\ref{eq: Path restriction}) then implies that $P_{n,f}$ approaches $P_0$ at rate $\frac{1}{\sqrt{n}}$, so the difference between the two distributions is on the same order of magnitude as sampling uncertainty.
This ensures that $P_{n,f}$ and $P_{0}$ remain ``close'' in a statistical sense even in large samples.

While the score $f$ controls the distribution of the data, our interest lies in the structural parameter $\theta^*$.  Identifying information about $\theta^*$ comes from the fact that not all elements of $T(P_0)$ are consistent with a given $\theta^*$. Specifically, the scaled sample average of the moments has (asymptotic) mean zero at $\theta^*$ under $P_{n,f}$ if and only if  $\mathbb{E}_{P_0}\left[f(X)\phi(X,\theta^*)\right]=0$. Correspondingly, the sub-space of scores consistent with $\theta^*$ is
$$ T_{\theta^*}(P_0)=\left\{f\in T(P_0):\mathbb{E}_{P_0}\left[f(X)\phi(X,\theta^*)\right]=0 \right\}.$$
This space is typically infinite-dimensional, in keeping with the semi-parametric nature of the GMM model.
 We are now equipped to define the  finite sample statistical experiment.
\begin{definition}
The finite sample experiment for sample size $n$, $\mathcal{E}_{n}^{*}$, corresponds to observing an
i.i.d. sample of random variables
$X_{i},i=1,...,n,$  distributed according to $P_{n,f}$, with parameter space $\left\{(\theta^*,f):\theta^*\in\Theta_0,f\in T_{\theta^*}(P_0)\right\}$.
\end{definition}

\begin{figure}
\includegraphics[scale=0.4]{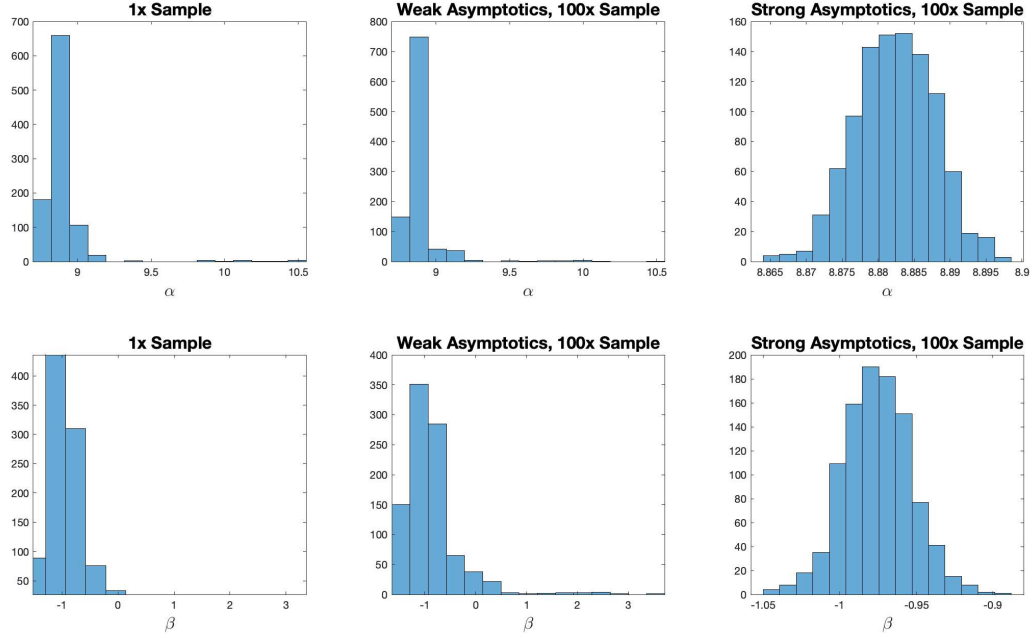}
\caption{Finite sample distribution of GMM estimator in simulations calibrated to Graddy (1995) data, along with weak- and strong-asymptotic large-sample distributions.  Based on 1000 simulation draws, dropping top and bottom 1\% of draws in the left panel for visibility.\label{fig: q75 approximation}}
\end{figure}

\paragraph{Example: Quantile IV, continued}
A variety of different distributions $P_0$ give rise to non-trivial identified sets $\Theta_0$ in this example.  Correspondingly, there are many ways weak identification may arise. For example, suppose that the first element of $Z$ is a constant, while the remaining elements of $Z$  can be written as $U Z^*$, for $U$ a mean-zero random variable independent of $(Y,W,Z^*)$ and $Z^*$ a potentially informative, but unobserved, instrument.
The last $k-1$ elements of $\mathbb{E}_{P_0}[\phi\left(X,\theta\right)]$ are thus identically zero on $\Theta,$ while the first element of $\mathbb{E}_{P_0}[\phi\left(X,\theta\right)]$ is zero if and only if $\alpha$ is equal to the $\tau$-th quantile of $Y-W'\beta$ under $P_0$, $q_{P_0,\tau}(Y-W'\beta)$. Hence the identified set under $P_0$ is
$\Theta_0=\{\theta=(\alpha,\beta)\in\Theta: \alpha=q_{P_0,\tau}(Y-W'\beta)\}$.

To illustrate the utility of weak identification asymptotic approximations, we return to our simulations calibrated to Graddy (1995). The second column of Figure \ref{fig: q75 approximation} shows the distribution of GMM estimators when we (a) increase the sample size by a factor of 100 and (b) model the finite-sample distribution $P_{n,f}$ as converging (in the sense of (\ref{eq: Path restriction})) to the distribution $P_0$ discussed above, with score $f$ chosen to match the distribution at the original sample size. For comparison, the third column shows the distribution under strong-identification asymptotics, which hold the data generating process fixed as $n$ grows. As Figure \ref{fig: q75 approximation} makes clear, weak-identification asymptotics preserve the asymmetric and heavy-tailed behavior seen in the data-calibrated design.  By contrast, under strong-identification asymptotics the distribution appears approximately normal.  This is expected given standard asymptotic results, but is a much worse approximation to the small-sample distribution.  $\Box$

\subsection{Asymptotic Representation Theorem}

This section shows that to construct asymptotically optimal decision rules for weakly identified GMM, it suffices to derive optimal decision rules in a limit experiment.
This limit experiment corresponds to observing a Gaussian process  $g(\cdot)$ with unknown mean function $m(\cdot)$ and known covariance function $\Sigma(\cdot,\cdot)$, where $\theta^*$ satisfies $m(\theta^*)=0$.
Intuitively, $g(\cdot)$ corresponds to the scaled sample average of the moments, since as we discuss below,  under $P_{n,f}$
\begin{align}\label{eq: convergence of process}
\frac{1}{\sqrt{n}}\sum_{i=1}^n\phi(X_i,\cdot)=g_n(\cdot)\Rightarrow g(\cdot)\sim\mathcal{GP}(m,\Sigma),
\end{align}
on $\Theta_0$, where $\Rightarrow$ denotes weak convergence as a process, $m(\cdot)=\mathbb{E}_{P_0}[f(X)\phi(X,\cdot)]$, $\Sigma(\theta_{1},\theta_{2})=\mathbb{E}_{P_0}[\phi(X,\theta_{1})\phi(X,\theta_{2})']$, and $\Sigma$ is consistently estimable.  We assume throughout that $\Sigma$ is continuous in both arguments.\footnote{Together with compactness of $\Theta_0$, this implies that we can take the stochastic process $g$ to be everywhere continuous almost surely. See Lemma 1.3.1 of Adler and Taylor (2007). }

To derive the limit experiment, we first discuss the parameter space for the mean function $m(\cdot)$.
Define a linear transformation mapping $f\in T(P_0)$ to functions  $m(\cdot)$:
\begin{equation}\label{eq:isomorphism}
m(\cdot)=\mathbb{E}_{P_0}[f(X)\phi(X,\cdot)].
\end{equation}
Let $\mathcal{H}$ be the image of $T(P_0)$ under this transformation. Lemma \ref{lem: on correspondence of rkhs} in the Appendix shows that $\mathcal{H}$ is the Reproducing Kernel Hilbert Space
(RKHS)  associated with the covariance function $\Sigma$.
Let $\mathcal{H}_{\theta^*}=\left\{m\in\mathcal H:m(\theta^*)=0\right\}$ denote the subset of $\mathcal{H}$ with a zero at $\theta^*$.

\begin{definition}

The Gaussian process experiment $\mathcal{E}_{GP}^{*}$  corresponds to observing a Gaussian
process $g(\cdot)\sim\mathcal{GP}(m(\cdot),\Sigma)$ with known covariance
function $\Sigma(\cdot,\cdot)$, unknown mean function $m$, and parameter space
$\left\{(\theta^*,m):\theta^*\in\Theta_0,m\in \mathcal{H}_{\theta^*}\right\}$.

\end{definition}

To compare $\mathcal{E}_{GP}^{*}$ to $\mathcal{E}_{n}^{*}$, we need to relate their parameter spaces. A challenge here is that the space of scores $f$ in the experiment $\mathcal{E}_{n}^{*}$ is larger than the space of means $m$ in $\mathcal{E}_{GP}^{*}$. To address this, we decompose each score as $f=f^*+f^\bot$ where $f^*$ is isomorphic to $m,$ while $f^\perp$ reflects aspects of the data generating process which are irrelevant to the large-sample behavior of the moments, and so can safely be ignored when constructing decision rules.
Specifically, denote the null space of the linear operator (\ref{eq:isomorphism}) by $\mathcal{H}^\perp$. By definition $\mathcal{H}^\perp$ is a linear subspace of $T(P_0)$.  Let $\mathcal{H}^*$ be the orthogonal complement of $\mathcal{H}^\perp$.\footnote{The proof of Lemma \ref{lem: on correspondence of rkhs} shows that $\mathcal{H}^*$ is the completion of the
space spanned by scores of the form $f(X)=\sum_{j=1}^{s}\phi(X,\theta_{j})'b_{j}$ in
$L_{2}({P_0})$.} For any score $f\in T(P_0),$ denote by $f^*$  and $f^{\bot}$ its projections onto $\mathcal{H}^{*}$ and $\mathcal{H}^\bot$ respectively.  By definition $f^\perp$ is orthogonal to the GMM moment condition, so two scores imply the same mean function if and only if they have the same projection $f^*$ onto $\mathcal{H}^*$.  We can re-write the parameter space for $\mathcal{E}_{n}^*$ as a Cartesian product
$$\{\theta^*\in \Theta_0,f=(f^*,f^\bot)\in T_{\theta^*}(P_0)\}=\{\theta^*\in \Theta_0,f^*\in \mathcal{H}^*_{\theta^*}\}\times\{f^\bot\in \mathcal{H}^\bot\}$$
 for $\mathcal{H}^{*}_{\theta^*}=\left\{f^*\in\mathcal{H}^*:\mathbb{E}_{P_0}[f^*(X)\phi(X,\theta^*)]=0\right\}$.
For fixed $f^\perp$, however, (\ref{eq:isomorphism}) defines an isomorphism between the parameter space in $\mathcal{E}_{n}^*$ and that in $\mathcal{E}_{GP}^*$, so following the literature on limits of experiments (c.f. Le Cam 1986) we can compare these experiments in terms of the attainable risk (expected loss) functions.

 \begin{theorem}\label{thm: GP representation theorem}
 Consider any sequence of statistics $S_{n}$ which has a limit distribution
under $\mathcal{E}_{n}^{*},$ in the sense that under $P_{n,f}$ for any $f\in T(P_0),$
$
S_{n}\left(X_{1},...,X_{n}\right)\Rightarrow S_{f}
$
as $n\to\infty.$
Assume there exists a complete separable set \textbf{$\mathbb{S}_{0}$}
such that $Pr\left\{S_{f}\in\mathbb{S}_{0}\right\}=1$ for all $f\in T(P_0)$, and that the loss function $L$  is continuous in its first argument. For any $f^\bot\in \mathcal{H}^\bot$ there exists in  $\mathcal{E}_{GP}^*$ a (possibly randomized) statistic
$
S
$
such that, defining $f=f^*+f^\perp$ and $m(\cdot)=\mathbb{E}_{P_0}[f(X)\phi(X,\cdot)]$,
$$ \liminf_{n\to\infty}\mathbb{E}_{P_{n,f}}[L(S_n,\theta^*)]\ge \mathbb{E}_{m}[L(S,\theta^*)] \text{ for all }\theta^*\in\Theta_0,f^*\in\mathcal{H}^*_{\theta^*},$$
where $\mathbb{E}_{m}$ is the expectation in $\mathcal{E}_{GP}^*$ under mean function $m$.
 \end{theorem}

Theorem \ref{thm: GP representation theorem} establishes that the attainable risk functions in $\mathcal{E}_{GP}^*$  lower-bound the attainable asymptotic risk functions in $\mathcal{E}_{n}^{*}$.
 Thus, if a sequence of decision rules $S_n$ has risk converging to an optimal risk function in $\mathcal{E}_{GP}^{*}$, it must be asymptotically optimal.
 The proof of this result builds on van der Vaart (1991), who derives the limit experiment for inference on $f$.  In our setting, however, $f^\perp$ is a nuisance parameter that neither interacts with the parameter of interest $\theta^*$ nor enters the loss function. Thus, to derive optimal procedures it suffices to study optimality holding $f^\perp$ fixed, similar to the ``slicing'' argument of Hirano and Porter (2009).

Theorem \ref{thm: GP representation theorem} gives a criterion which may be checked to verify asymptotic optimality.  In many cases, a plug-in approach further suggests the form of an asymptotically optimal rule.
Suppose we know an optimal decision rule $S=s(g,\Sigma)$ in  $\mathcal{E}_{GP}^{*}$, where we now make dependence on the covariance function explicit.\footnote{All of our results allow for randomized decision rules, e.g. $S=s(g(\cdot),\Sigma,U)$ with $U\sim U[0,1]$ independent of the data, but for simplicity we suppress randomization in our notation.}
If a uniform central limit theorem holds under $P_0$ and we have a consistent estimator $\widehat\Sigma$ for $\Sigma$  (e.g. the sample covariance function $\widehat\Sigma(\theta,\tilde\theta)=\widehat{Cov}(\phi(X_i;\theta),\phi(X_i;\tilde\theta))$), then Le Cam's third lemma implies that
the weak convergence (\ref{eq: convergence of process}) holds, while $\widehat\Sigma$ remains consistent under $P_{n,f}$. Provided $s(g,\Sigma)$ is almost-everywhere continuous in $(g(\cdot),\Sigma)$, the Continuous Mapping Theorem thus implies that
$S_n=s\left(g_n(\cdot),\widehat\Sigma\right)\Rightarrow s(g,\Sigma)=S$
under $P_{n,f},$ so the sequence of rules $S_n$ is asymptotically optimal so long as convergence in distribution implies convergence in expected loss (e.g. under uniform integrability conditions).


\paragraph{Special Case: Finite $\Theta$}

Suppose that
 $\Theta_{0}=\left\{ \theta_{1},...,\theta_{r}\right\} $ is finite.
In this case the Gaussian process experiment $\mathcal{E}^*_{GP}$ reduces to observing
the normal random vector $g=\left(g\left(\theta_{1}\right)',...,g\left(\theta_{r}\right)'\right)'\in\mathbb{R}^{r k}$
with unknown mean $m=\left(m\left(\theta_{1}\right)',...,m\left(\theta_{r}\right)'\right)'$
and known $rk\times rk$ variance matrix $\Sigma$. We assume for this example that $\Sigma$ has full rank, which implies that $\mathcal{H}=\mathbb{R}^{r k}$.  The parameter space in $\mathcal{E}_{GP}^{*}$ thus consists of $(\theta^*,m)$ pairs where the sub-vector of $m$ corresponding to $\theta^*$ is zero: $\left\{ \theta^{*}\in\Theta_{0},m\in\mathbb{R}^{r k}:m\left(\theta^{*}\right)=0\right\} $. Hence, the limit problem reduces to observing a collection of normal
random vectors and trying to infer which among them have mean zero.

Even for finite $\Theta_0$, the score $f$ in the finite-sample experiment
$\mathcal{E}_{n}^{*}$ is typically infinite-dimensional and controls all features
of the data distribution, for instance the skewness and kurtosis of
 $\phi\left(X,\theta\right)$. Asymptotically,
however,
only part of $f$ is important, namely, its projection
$f^{*}\left(X\right)=\phi\left(X\right)'\Sigma^{-1}\mathbb{E}_{P_{0}}\left[\phi\left(X\right)f\left(X\right)\right]$ on $\phi\left(X\right)=\left(\phi\left(X,\theta_{1}\right)',...,\phi\left(X,\theta_{r}\right)'\right)',$
where there is a one-to-one correspondence between scores $f^{*}$
and $m\in\mathcal{H}$. Theorem  \ref{thm: GP representation theorem} then
implies that (under regularity conditions) it is without loss, in terms of attainable asymptotic performance,
to limit attention to decision rules that depend on the data only
through the sample average moments $\frac{1}{\sqrt{n}}\sum\phi\left(X_{i}\right)$
and the estimated variance $\widehat{\Sigma}$. $\Box$

The idea of solving a limit problem in order to derive asymptotically optimal decision rules is of course not new -- see e.g. Le Cam (1986).  More recently, Mueller (2011) proposed an alternative approach to derive asymptotically optimal tests starting from weak convergence conditions like (\ref{eq: convergence of process}). Relative to the approach of Mueller (2011) applied to our setting, the benefits of  Theorem \ref{thm: GP representation theorem} are (i) to show that there is, in a sense, no asymptotic information loss from limiting attention to the sample average of the moments and (ii)  the ability to consider general decision problems in addition to tests.

\section{Priors for GMM Limit Problem}\label{section- prior}
The previous section showed that we can reduce the search for asymptotically optimal decision rules to a search for rules based on the Gaussian process
$
g(\cdot)\sim \mathcal{GP}(m,\Sigma),
$
with a known covariance function $\Sigma(\cdot,\cdot)$ and an unknown mean function $m$ such that $m(\theta^*)=0$.

While the Gaussian process limit experiment is much simpler than the original finite-sample GMM setting, it still has an infinite-dimensional parameter space.
Moreover minimizing risk at different parameter values generally produces different decision rules, so there do not exist uniformly best estimators or uniformly most powerful tests.
Instead, optimal decision rules trade off performance over different regions of the parameter space. In this paper we focus on Bayes decision rules, which make this tradeoff explicit through the prior.
The Bayes decision rule  for the prior $\pi$ minimizes $\pi$-weighted average risk,  $\min_{\tilde{s}} \int \mathbb{E}_{\theta^*,m}[L(\tilde{s}(g),\theta^*)]d\pi(\theta^*,m)$.

Bayes decision rules are closely linked to optimality, and in particular admissibility, in the limit problem.
A decision rule $s(g)$ in the experiment $\mathcal{E}_{GP}^*$ is admissible if there exists no rule $\tilde{s}(g)$ with weakly lower risk for all parameter values, and strictly lower risk for some.  A class $\mathcal{S}$ of rules is complete if it contains all admissible rules.
For convex loss functions, a result in Brown (1986) implies that pointwise limits of Bayes decision rules are a complete class.

\begin{theorem}\label{Thm: Brown admissibility} (Brown, 1986) Suppose that $\mathcal{A}$ is closed, with $\mathcal{A}\subseteq\mathbb{R}^{d}$ for some $d$, that $L\left(a,\theta\right)$ is continuous and strictly
convex in $a$ for every $\theta$, and that either $\mathcal{A}$ is bounded or $\lim_{\|a\|\to\infty}L(a,\theta)=\infty$.  Then for every admissible decision rule $s$ in $\mathcal{E}_{GP}^*$ there exists a sequence of priors $\pi_r$ and corresponding Bayes decision rules $s_{\pi_r}$,
$$
\int E_{\theta^*,m}[L(s_{\pi_r}(g),\theta)]d\pi_r(\theta^*,m)=\min_{\tilde{s}} \int E_{\theta^*,m}[L(\tilde{s}(g),\theta)]d\pi_r(\theta^*,m),
$$
such that $s_{\pi_r}(g)\to s(g)$ as $r\to \infty$ for almost every $g$.
\end{theorem}
Related results hold for settings with non-convex loss, e.g. in testing problems.  For instance,  Strasser (1985, Theorem 47.9) shows that the weak limits of Bayes decision rules form an essentially complete class in settings with bounded loss.

Even focusing on Bayes decision rules, it remains to chose a prior $\pi(\theta^*,m)$ and to derive the  resulting optimal rule.  The infinite-dimensional nature of the limit experiment makes it difficult to specify subjective priors, so in this section we propose that researchers use a subjective prior for the structural parameter $\theta^*$ together with a default prior on an infinite-dimensional nuisance parameter.  While these default priors are motivated by conjugacy and invariance arguments, it turns out that all priors in the class we consider deliver the quasi-Bayes approach suggested by Chernozhukov and Hong (2003) as a limiting case.
\subsection{A Class of Priors for the Limit Problem}

We derive our class of priors in three steps.  First, we re-parameterize the limit experiment to separate the structural parameter $\theta^*$ from an infinite-dimensional nuisance parameter.  Second, for tractability we consider independent priors on the two components.  Third, we use a conjugate Gaussian prior on the nuisance parameter, and show that a natural invariance property for default priors greatly restricts the choice of prior covariance.  In particular, invariance generically limits attention to what we term proportional priors, which are indexed by a scalar variance parameter.  Taking this variance parameter to infinity yields the quasi-Bayes approach.

\subsubsection{Reparameterization}

The parameter space $\left\{(\theta^*,m):\theta^*\in\Theta_0,m\in \mathcal{H}_{\theta^*}\right\}$ in $\mathcal{E}_{GP}^*$ requires that for fixed $\theta^*,$ the mean function $m$ must lie in the linear subspace $\mathcal{H}_{\theta^*}$, so the parameter space for the infinite-dimensional parameter $m$ depends on $\theta^*$. To simplify the analysis, we re-parameterize the model to disentangle $\theta^*$ from the nuisance parameter.

Denote by $\mathcal{C}$ the space of continuous  functions from $\Theta_0$ to $\mathbb{R}^k$, and let $A$ be any linear functional $A:\mathcal{C}\to\mathbb{R}^k$ such that $Cov(A(g),g(\theta))$ is nonsingular for all $\theta$, where we assume such a functional $A$ exists.
\begin{lemma}\label{lem: reparametrization}
Let $\xi=A(g),$ $h(\cdot)=g(\cdot)-Cov(g(\cdot),\xi)Var(\xi)^{-1}\xi$, and $\mu(\cdot)=m(\cdot)-Cov(g(\cdot),\xi)Var(\xi)^{-1}A(m)$.
Then there exist  one-to-one correspondences (i) between $\left\{(\theta^*,m):\theta^*\in\Theta_0,m\in \mathcal{H}_{\theta^*}\right\}$ and $(\theta^*,\mu)\in \Theta_0\times \mathcal{H}_\mu$, and (ii) between $g(\cdot)$ and $(\xi,h(\cdot))$ such that:
\begin{itemize}
\item[(a)] The vector $\xi\sim N(\nu(\theta^*,\mu),\Sigma_\xi)$  and process $h(\cdot)\sim \mathcal{GP}(\mu,\widetilde\Sigma)$ are independent, where $\nu$, $\Sigma_\xi$ ,and $\widetilde\Sigma$ are known transformations of $(\Sigma,A)$.
\item[(b)] $\mathcal{H}_\mu$ is the RKHS generated by  covariance function $\widetilde\Sigma$.
\end{itemize}
\end{lemma}
Lemma \ref{lem: reparametrization} re-parameterizes the model in terms of the structural parameter $\theta^*$ and a functional nuisance parameter $\mu$, where the parameter space for $(\theta^*,\mu)$ is a Cartesian product and the parameter space for $\mu$ is linear.  Intuitively, each $\mathcal{H}_{\theta^*}$ imposes the $k$-dimensional linear restriction $m(\theta^*)=0$, while the reparameterization expresses these $\theta^*$-specific restrictions as restrictions on a common $k$-dimensional functional $A(m),$ leaving the remainder of $m,$ described by $\mu,$ unrestricted.
Lemma \ref{lem: reparametrization}  also establishes a corresponding decomposition of the observed process $g(\cdot)$ into components $(\xi,h(\cdot)),$ where the distribution of $h$ depends only on $\mu$.

\paragraph{Special Case: Point Evaluation} Suppose $A$ is the point evaluation functional at some value $\theta_0$, so $\xi=A(g)=g(\theta_0)$.  Andrews and Mikusheva (2016) show that
  \begin{equation}\label{eq: def of h}
h(\cdot)=g(\cdot)-{\Sigma}(\cdot,\theta_0){\Sigma}(\theta_0,\theta_0)^{-1}g(\theta_0),
\end{equation}
is a Gaussian process  independent of $\xi\sim N(m(\theta_0), \Sigma (\theta_0,\theta_0))$, while Lemma \ref{lem: reparametrization} establishes that if $\Sigma(\cdot,\theta_0)$ is everywhere full rank, knowledge of the functional parameter $$\mu(\cdot)=\mathbb{E}[h(\cdot)]=m(\cdot)-{\Sigma}(\cdot,\theta_0){\Sigma}(\theta_0,\theta_0)^{-1}m(\theta_0)$$
and the structural parameter $\theta^*$ suffices to reconstruct the mean function $m$. $\Box$

Analogous reparameterizations can be constructed for  other linear functionals $A$, and we obtain a different parameterization for each such functional.  Since $A$ does not matter for the procedure we ultimately recommend, we do not discuss how to choose it.

\paragraph{Special Case: Finite $\Theta$, continued}

When $\Theta$ is finite, any $k$-dimensional linear functional $A$ with $A(0)=0$ corresponds to a $k\times r k$ matrix.
In this case $\xi=Ag$ is a $k\times 1$ Gaussian vector, while $h=\left(h\left(\theta_{1}\right)',...,h\left(\theta_{r}\right)'\right)'=\left(I-\Sigma A'\left(A\Sigma A'\right)^{-1}A\right)g$ is a Gaussian vector with mean $\mu=\left(I-\Sigma A'\left(A\Sigma A'\right)^{-1}A\right)m$ and rank $(q-1)k$ covariance matrix $\widetilde \Sigma= \Sigma-\Sigma A'\left(A\Sigma A'\right)^{-1}A\Sigma$.  In this context our assumption that $Cov(Ag,g(\theta))$ is nonsingular for all $\theta$ is equivalent to requiring that all $r$ of the  $k\times k$ blocks in the $k\times rk$ matrix $A\Sigma$ have full rank. Since $\Sigma$ has full rank, such $A$ always exists.
Lemma \ref{lem: reparametrization} then states that the transformation from $\left(\theta^{*},m\right)$
to $\left(\theta^{*},\mu\right)$ is one-to-one, and hence a reparameterization.
The original parameter space for $m$ is the subset of $\mathbb{R}^{rk}$ where at least one of $k$-dimensional component is exactly zero.  This space is not linear, or even convex, as convex combinations of two mean vectors with zeros at different $\theta$'s need not be zero anywhere. By contrast, the parameter space for $\left(\theta^{*},\mu\right)$ is the
Cartesian product $\Theta_{0}\times\mathcal{H}_{\mu}$, where $\mathcal{H}_{\mu}=Span\left\{ \widetilde{\Sigma}\right\} $ is a $(r-1)k$ dimensional linear subspace of $\mathbb{R}^{rk}$.  $\Box$

\subsubsection{Prior Independence}

The likelihood function $\ell\left(\mu,\theta^*;g\right)$ based on the observed data $g(\cdot)$ factors as\footnote{All Gaussian processes with covariance function $\Sigma$ and mean functions in $\mathcal{H}$ are mutually absolutely continuous, so we can define the likelihood with respect to any base measure in this class.}
$$
\ell\left(\mu,\theta^*;g\right)=\ell\left(\mu,\theta^*;\xi\right)\ell\left(\mu;h\right),
$$
where $\ell\left(\mu,\theta^*;\xi\right)$ and $\ell\left(\mu;h\right)$ are the likelihood functions based on $\xi$ and $h$, with the latter depending only on $\mu$ and not on $\theta^*$.
Since the loss function depends only on $\theta^*$, to derive Bayes decision rules it suffices to construct the marginal posterior distribution for $\theta^*$.
For analytical tractability we consider independent priors $\pi(\theta^*)\pi(\mu)$ on $\theta^*$ and $\mu$. Under such priors the marginal  posterior for $\theta^*$ simplifies to
\begin{align}\label{eq: posterior}
\pi(\theta^*|g)=\frac{\pi(\theta^*)\ell^{*}\left(\theta^*\right)}{\int \pi(\theta)\ell^{*}\left(\theta\right)d\theta}, \text{ ~for~ }\ell^{*}\left(\theta\right)=\int \ell\left(\mu,\theta;\xi\right)d\pi\left(\mu|h\right),
\end{align}
where
$\pi(\mu|h)$ denotes the posterior for $\mu$ given $h$.

\subsubsection{Conjugacy and Invariance}

The posterior (\ref{eq: posterior}) depends on the data $(\xi,h(\cdot))$, the prior  $\pi(\theta^*)$ on the structural parameter, and the prior $\pi(\mu)$ on the nuisance parameter. Researchers may have informative priors about $\theta^*$ in a given application, but seem unlikely to have strong priors about $\mu$.  Moreover, the infinite-dimensional nature of $\mu$ makes it challenging to specify a subjective prior even when the researcher has prior information. We thus leave the prior on $\theta^*$ free to be selected by the researcher, while seeking a default recommendation for $\pi(\mu)$.

In our search for a default $\pi(\mu)$ we restrict attention to Gaussian process priors  $\mu\sim \mathcal{GP}(0,\Omega)$, for $\Omega(\cdot,\cdot)$ a continuous covariance function. This allows us to exploit conjugacy results, greatly simplifying computation of the integrated likelihood $\ell^*(\theta).$\footnote{For Gaussian  $\pi(\mu)$,  $\ell^{*}\left(\theta^*\right)$ corresponds to a Gaussian likelihood for observation $\xi$ with mean given by  the best linear predictor of $\xi$ based on $h$ under the prior. The solution to this linear prediction problem is obtained in Parzen (1962), and details appear in the proof of Theorem \ref{thm: on invariant prior}, stated below.}
Even with this restriction, the space
of potential covariance functions $\Omega$ is enormous, and it is challenging to directly evaluate
whether or not a given prior covariance is reasonable.
To derive default priors, we thus take a different approach, and ask what choices of
prior covariance $\Omega$ lead to decision rules with desirable properties.

We want a default $\pi(\mu)$ to imply reasonable decision rules when combined with many different choices of $\pi(\theta^*)$.
To this end, we impose that
 if a researcher rules out some parameter values ex-ante, the implied Bayes decision rules should not depend on the behavior of the moments at the excluded parameter values.
Formally, we require that for priors $\pi(\theta^*)$ with restricted support $\widetilde\Theta \subset\Theta_0,$ Bayes decision rules based on the prior $\pi(\theta^*)\pi(\mu)$ should depend on the data only through $\xi$ and the restriction of $h$ to $\widetilde\Theta$.
 For this invariance property to hold for all possible priors $\pi(\theta^*)$ and all loss functions $L(a,\theta^*)$, however, it must be that $\ell^*(\theta)$ depends on the data only through $(\xi,h(\theta))$ for all $\theta\in\Theta_0$.\footnote{See  Appendix B for details, and a formal invariance argument.
}
This restriction dramatically narrows the class of candidate covariance functions $\Omega$.

\begin{theorem}\label{thm: on invariant prior}
Assume the covariance function $\Omega$ is continuous.
\begin{itemize}
\item[(a)]
For all $\theta^*\in \Theta_0$ such that $\widetilde\Sigma(\theta^*,\theta^*)$ and $\Omega(\theta^*,\theta^*)$ have full rank, the integrated likelihood $\ell^{*}\left(\theta^*\right)$  depends on the data only through $(\xi, h(\theta^*))$ if and only if
\begin{align}\label{eq: condition on omega}
\Omega(\theta^*,\theta^*)^{-1}\Omega(\theta^*,\theta)=\widetilde\Sigma(\theta^*,\theta^*)^{-1}\widetilde\Sigma(\theta^*,\theta) \mbox{    for all  } \theta\in \Theta_0.
\end{align}

\item[(b)]
Assume, further, that for some  $\theta_0\in\Theta_0$ such that $\widetilde\Sigma(\theta_0,\theta_0)$ is full rank, there does not exist a non-trivial (non-empty, but strictly smaller than $\mathbb{R}^k$) linear subspace $V\subseteq\mathbb{R}^k$ that is invariant for the whole family of symmetric matrices\footnote{A linear subspace $V\subseteq\mathbb{R}^k$ is invariant for a linear operator $L$ if for any $v\in V$ we have $Lv\in V$. Invariant sub-spaces for a symmetric matrix $L$ are the sub-spaces spanned by subsets of its eigenvectors.}
 $$\mathcal{D}=\left\{D(\theta)=R(\theta_0,\theta)R(\theta_0,\theta)', \theta\in \Theta_0 : \det(\widetilde{\Sigma}(\theta,\theta))>0 \right\},$$
where $R(\theta_0,\theta)=\widetilde{\Sigma}(\theta_0,\theta_0)^{-1/2}\widetilde{\Sigma}(\theta_0,\theta) \widetilde{\Sigma}(\theta,\theta)^{-1/2}$ is a correlation function. Then condition (\ref{eq: condition on omega}) is equivalent to $\Omega(\cdot,\cdot)=\lambda\widetilde\Sigma(\cdot,\cdot)$ for some $\lambda>0$.
\end{itemize}
\end{theorem}

Two positive-definite matrices share an invariant subspace if and only if several eigenvectors of one matrix span the same sub-space as several eigenvectors of the other.  Generically (that is, everywhere but  a nowhere-dense subset) two positive-definite matrices share no non-trivial invariant subspace.  For the condition in part (b) of Theorem \ref{thm: on invariant prior}  to fail requires something still stronger, namely that the same subspace be invariant for a whole family of matrices indexed by $\theta$. Thus, invariance generically reduces the class of candidate priors to the one dimensional family $\Omega(\cdot,\cdot)=\lambda\widetilde{\Sigma}(\cdot,\cdot)$.

\paragraph{Special Case: Finite $\Theta$, continued}

Recall that in this case $\mathcal{H}_{\mu}=Span\left\{\widetilde{\Sigma}\right\}$ is a $(r-1)k$ dimensional linear subspace of $\mathbb{R}^{r k}$. Hence, $h\sim N\left(\mu,\widetilde{\Sigma}\right)$ has support $\mathcal{H}_{\mu}$, and the inverse of $\widetilde{\Sigma}$ is well defined on $\mathcal{H}_{\mu}$.
Consider a Gaussian prior $\mu\sim N(0,\Omega)$ with support $\mathcal{H}_\mu$.  This implies that $\Omega$ is also rank $(r-1)k$, with a well-defined inverse on $\mathcal{H}_{\mu}$. The posterior mean for $\mu$ given $h$ is thus $\left(\widetilde{\Sigma}^{-1}+\Omega^{-1}\right)^{-1}\widetilde{\Sigma}^{-1}h$.

The proof of Theorem \ref{thm: on invariant prior} establishes that the integrated likelihood
$\ell^{*}\left(\theta^{*}\right)$ depends on the data through $\left(\xi,h\left(\theta^{*}\right)\right)$
if and only if the posterior mean for $\mu\left(\theta^{*}\right)$
given $h$ depends only on $h\left(\theta^{*}\right).$ Given the formula for the posterior mean, the matrix $\left(\widetilde{\Sigma}^{-1}+\Omega^{-1}\right)^{-1}\widetilde{\Sigma}^{-1}$
must therefore be block-diagonal, with $r$ blocks of size $k\times k$.  We then show that, generically in the space of covariance
matrices $\widetilde{\Sigma}$, $\left(\widetilde{\Sigma}^{-1}+\Omega^{-1}\right)^{-1}\widetilde{\Sigma}^{-1}$ is block-diagonal if and only if $\Omega$ proportional to $\widetilde{\Sigma}$.
$\Box$

While the condition in Theorem \ref{thm: on invariant prior}(b) holds generically, it can fail in cases where the moments have special structure.  For instance, suppose a researcher forms moments based on two independent datasets, where one dataset is used to form the first group of moments while the other is used for the rest.  In this case $\widetilde\Sigma$ will be block-diagonal, and will imply two orthogonal invariant sub-spaces that are common across all $\theta$. If these are the only nontrivial invariant subspaces, the family of $\Omega$ satisfying condition (\ref{eq: condition on omega})  is two-dimensional, allowing a researcher to put different coefficients of proportionality on two invariant sub-spaces.

\subsection{Proportional Priors and Quasi-Likelihood}

Motivated by Theorem \ref{thm: on invariant prior}, we focus on proportional prior covariance functions, $\Omega(\cdot,\cdot)=\lambda\widetilde\Sigma(\cdot,\cdot)$.\footnote{Similar to e.g. Jeffreys prior, proportional priors depend on the data generating process and so violate the likelihood principal.} In this case
\begin{align*}
\ell^{*}\left(\theta\right)=\ell( \theta;  g,\Sigma,\lambda)=\left|\Lambda(\theta)\right|^{-\frac{1}{2}}\cdot \exp\left(-\frac{1}{2}u(\theta)'\Lambda(\theta)^{-1}u(\theta) \right),
\end{align*}
for
$
u(\theta)=\frac{\lambda}{1+\lambda}\psi\left(\theta\right)^{-1} g\left(\theta\right)+\frac{1}{1+\lambda}\xi
$,
$
\Lambda(\theta)=\frac{\lambda}{1+\lambda}\left[\psi\left(\theta\right)^{-1}\right] \Sigma\left(\theta,\theta\right)\left[\psi\left(\theta\right)^{-1} \right]'+\frac{1}{1+\lambda}Var(\xi),
$
where the $k\times k$-matrix valued function $\psi (\cdot)=Cov(g(\cdot),\xi)Var(\xi)^{-1}$ depends on  $A$ and $\Sigma$. Hence, for proportional priors the posterior distribution (\ref{eq: posterior}) takes a simple form.

The constant of proportionality $\lambda$ controls the strength of identification under the prior.  When $\lambda=0$ the prior implies that the mean function $m$ is zero with probability one, so nothing can be learned from $g$ and the posterior on $\theta^*$ is equal to the prior.  By contrast, under the diffuse ($\lambda\to\infty$) limit, the prior variance of $m$ diverges.
In this case
{\small
\begin{align}\label{eq: Quasi-likelihood}
\lim_{\lambda\to\infty}\ell( \theta;  g,\Sigma,\lambda)=\left|\psi(\theta)\right|\cdot\left|\Sigma(\theta,\theta)\right|^{-\frac{1}{2}}\cdot \exp\left(-\frac{1}{2}g(\theta)'\Sigma(\theta,\theta)^{-1}g(\theta) \right).
\end{align}
}Hence, as $\lambda\to\infty$, $\ell^{*}\left(\theta\right)$ converges to a transformation of the continuously updating GMM objective function, multiplied by factors that do not depend on $g$ and so may be absorbed into the prior.  Note, in particular, that $A$ enters (\ref{eq: Quasi-likelihood}) only through $\left|\psi(\theta)\right|$, so as $\lambda\to\infty$ our default priors deliver posteriors of the same form for all choices of $A$.

The quasi-likelihood (\ref{eq: Quasi-likelihood}) was used by Chernozhukov and Hong (2003) as part of their quasi-Bayes approach, discussed further in the next section.  Motivated by the simplifications obtained by taking $\lambda\to\infty$ (e.g. the elimination of dependence on $A$) together with other desirable properties discussed in the following sections, we recommend using this limiting quasi-likelihood, absorbing $\left|\psi(\theta)\right|$ and $\left|\Sigma(\theta,\theta)\right|^{-\frac{1}{2}}$ into the prior to obtain a quasi-posterior of the form (\ref{eq: posterior}) with $\ell^*(\theta)=\exp(-\frac{1}{2}Q(\theta))$ for $Q(\theta)=g\left(\theta\right)'{\Sigma} \left(\theta,\theta\right)^{-1}g\left(\theta\right).$

While (\ref{eq: Quasi-likelihood}) resembles a Gaussian likelihood, under weak identification the function $Q(\theta)$  will often be far from quadratic, implying highly non-normal quasi-posteriors for the structural parameter $\theta$.  In strongly-identified settings,  by contrast, the data rule out values of $\theta$ outside a small neighborhood of $\theta^*$, allowing Taylor expansion arguments to establish that the function $Q(\theta)$ is approximately quadratic in $\theta$ and the quasi-posterior for $\theta$ is approximately normal.  Intuitively, the Gaussian form of (\ref{eq: Quasi-likelihood}) reflects that the moment function, evaluated at a given point, is asymptotically normal, but under weak identification this does not imply normality of the quasi-posterior for the structural parameter.

\section{Decision Rules Based on the Quasi-Posterior}\label{sec: quasi-posterior rules}

The last section derived a class of priors for the limit problem, and showed that they imply the quasi-likelihood (\ref{eq: Quasi-likelihood}) as a limiting case.  This section discusses how to use this quasi-likelihood to construct decision rules.  We first consider point estimation and other decision problems with no constraint on the class of rules, and then turn to statistical tests, which impose frequentist size control as a side condition.

\subsection{Estimation}

We first consider quasi-Bayes estimates, or other unconstrained decision rules, which solve the weighted average risk minimization problem over the class of all decision rules.  To minimize weighted average risk it suffices to minimize the posterior expected loss for each realization of $g$, taking
\begin{align}\label{eq: Bayes decision rule- limit}
s(g)\in\arg\min_{a\in\mathcal{A}}\int_{\Theta_0}L(a,\theta)d\pi(\theta|g)=\arg\min_{a\in\mathcal{A}}\frac{\int_{\Theta_0}L(a,\theta)\pi(\theta) \exp\left\{-\frac{1}{2}Q\left(\theta\right)\right\}d\theta}{\int_{\Theta_0} \pi(\theta)\exp\left\{-\frac{1}{2}Q\left(\theta\right)\right\}d\theta}.
\end{align}
For instance, under squared-error loss $L(a,\theta)=\|a-\theta\|_2^2$, equation (\ref{eq: Bayes decision rule- limit}) delivers the quasi-posterior mean.  If $\theta$ is scalar and we instead consider the check-function loss $L(a,\theta)=(\tau-\mathbb{I}\{\theta\le a\})(\theta-a)$, the optimal decision rule  corresponds to the $\tau$-th quasi-posterior quantile.
While the quasi-posterior $\pi(\theta|g)$ is not in general available in closed form, there is a large Bayesian computational literature on how to obtain posterior draws (e.g. via Markov chain Monte Carlo).  Given a sample of such draws, we can then approximate the optimal decision rule by minimizing the sample average loss.  Chernozhukov and Hong (2003) provide a detailed discussion of computation of quasi-Bayes decision rules (\ref{eq: Bayes decision rule- limit}).

Chernozhukov and Hong (2003) advocate the quasi-Bayes approach as a computational device for point-identified, strongly-identified settings where Bayesian techniques are more numerically tractable than minimization, and show that the resulting estimators are asymptotically efficient (again under strong identification). We have established that the same quasi-posterior arises as the limit of proper-prior Bayes posteriors even under weak identification.  Appendix C shows that under further regularity conditions, quasi-Bayes decision rules correspond to pointwise limits of proper-prior (and  admissible) Bayes decision rules, consistent with the conditions of Theorem \ref{Thm: Brown admissibility}.  Section \ref{sect - implementation} establishes the asymptotic properties for a feasible version of these rules.  Given the range of desirable properties established for these rules, we recommend the use of quasi-Bayes decision rules in settings where weak identification is a concern.

\paragraph{Bayesian Approaches to GMM}

Several other papers have justified quasi-Bayes decision rules  (\ref{eq: Bayes decision rule- limit}) from a Bayesian perspective.  Closest to our approach, Florens and Simoni (2019) consider Bayesian inference based on an asymptotic normal approximation to a transformation of the data, and obtain the quasi-likelihood (\ref{eq: Quasi-likelihood}) as a diffuse-prior limit.  Unlike our analysis, however, they specify a Gaussian process prior on the finite-sample density of the data $X$, rather than on the mean function in the limit experiment.  Earlier work by Kim (2002) obtained the same quasi-likelihood via maximum entropy arguments, while Gallant (2016) obtains it as a Bayesian likelihood based on a coarsened sigma-algebra.  Unlike our analysis,  none of these papers speak to questions of optimality.

Other authors have considered alternative Bayesian approaches for moment condition models that do not run through the quasi-likelihood (\ref{eq: Quasi-likelihood}).  Chamberlain and Imbens (2003) consider inference for just-identified moment condition models with discrete data, while Bornn et al. (2019) consider discrete data and potentially over-identified moment conditions.  Both procedures have a finite-sample Bayesian justification, unlike our approach.  Kitamura and Otsu (2011) and Shin (2015) consider Bayesian approaches based on Dirichlet process priors and exponential tilting arguments.  Finally, Lazar (2003) discusses posteriors formed using the empirical likelihood objective, and Schennach (2005) shows that a particular generalized empirical likelihood (GEL)-type objective function arises in the limit for a family of nonparametric priors.

It may be surprising that the quasi-likelihood (\ref{eq: Quasi-likelihood}) is simply a transformation of the limiting GMM objective function, since there are many other objectives that can be used to fit moment conditional models, particularly GEL approaches.  Guggenberger and Smith (2005) show, however, that under regularity conditions all GEL objective functions are first-order asymptotically equivalent under weak (as well as strong) identification.  Since continuously updating GMM is a member of the GEL family, this implies that all approaches in this family, including Bayesian empirical likelihood as proposed by Lazar (2003), are equivalent to (\ref{eq: Quasi-likelihood}) in the limit experiment.  Since the approach of Schennach (2005) is closely related to GEL, it is likewise equivalent to (\ref{eq: Quasi-likelihood}) in the limit problem.\footnote{The equivalence proofs in Guggenberger and Smith (2005) extend directly to this case.}

\subsection{Optimal Tests}\label{sec: tests}

Our results show desirable properties for quasi-Bayes decision rules,  but Bayesian inference does not in general control frequentist size or coverage under weak identification.  In particular, while Chernozhukov and Hong (2003) show that credible sets formed based on quasi-posterior quantiles have correct coverage when identification is strong, these sets can badly under-cover when identification is weak, consistent with well-known results on Bayesian inference for settings with identification failure (e.g. Moon and Schorfheide, 2012).
Our results are nonetheless useful for frequentist inference, since Theorem \ref{thm: GP representation theorem} implies that attainable asymptotic type-I and type-II error functions are bounded below by the corresponding functions in the limit experiment.  Hence, if we can characterize optimal tests in the limit experiment, these bound the attainable asymptotic power.

In previous work (Andrews and Mikusheva, 2016) we developed a general technique for constructing identification-robust tests based on a wide variety of test statistics.  Different choices of test statistic imply different power properties, however, and the optimal choice of statistic remained an open question.  Here, we characterize optimal tests.

Consider the problem of testing  the null $H_{0}:\theta^{*}=\theta_{0}$ against the composite alternative $H_{1}:\theta^{*}\neq\theta_{0}$. For this problem it is natural to let $A$ be the point evaluation functional at $\theta_0,$ $A\left(m\right)=m\left(\theta_{0}\right)$. Define $\mu$ and $h$ as in Lemma \ref{lem: reparametrization} and the point evaluation example.  There is no uniformly most powerful test in this setting, so we instead specify weights $\pi(\theta^*,\mu)$ over parameter values in the alternative.
We want to maximize weighted average power (WAP), subject to controlling the rejection probability under the null
$$
\max_{\varphi} \int \mathbb{E}_{\theta^*,\mu}[\varphi]d\pi(\theta^*,\mu)\text{ subject to } \mathbb{E}_{\theta_0,\mu}[\varphi]\le\alpha \text{ for all }\mu\in\mathcal{H}_{\mu},
$$
where $\varphi=0$ and $\varphi=1$ denote non-rejection and rejection of $H_0,$ respectively.
Unfortunately, the size constraint in this setting is challenging to work with, since it involves the infinite-dimensional nuisance parameter $\mu\in\mathcal{H}_{\mu}$.
If we strengthen the size constraint to require similarity,
\begin{equation}\label{eq: similarity}
\mathbb{E}_{\theta_0,\mu}[\varphi]=\alpha\text{ for all }\mu\in\mathcal{H}_{\mu},
\end{equation}
 this simplifies the problem.\footnote{WAP-optimal similar tests for the weak IV model have been studied by D. Andrews, Moreira, and Stock (2006), Moreira and Moreira (2019), and Moreira and Ridder (2019), among others.  Specialized to linear IV, the method we propose here is equivalent to a particular class of priors.} Theorem S2.1 in the supplement to Andrews and Mikusheva (2016) establishes that any test $\varphi$ satisfying (\ref{eq: similarity}) must be conditionally similar, with $\mathbb{E}_{\theta_0}[\varphi|h]=\alpha$ for almost every $h$, and shows how to construct conditionally similar tests. Together with the results above, this delivers WAP-optimal similar tests:

 \begin{theorem}\label{thm: testing}
Let $T=\frac{\int\ell^{*}\left(\theta\right)\pi\left(\theta\right)d\theta}{\ell^{*}\left(\theta_{0}\right)}$, where $\ell^*(\cdot)$ is defined in (\ref{eq: posterior}).  Let $c_{\alpha}(h)$ be $1-\alpha$ quantile of the conditional distribution of $T$ given $h$ under $\theta_0$. Provided the distribution of $T$ given $h$ is almost surely continuous, the test $\varphi^*(\theta_0)=\mathbb{I}\{T>c_{\alpha}(h)\}$ is similar.  Further, $\varphi^*(\theta_0)$
maximizes $\pi\left(\theta,\mu\right)=\pi\left(\theta\right)\pi\left(\mu\right)$-weighted average power
over the class of similar tests, in the sense that for any other test
$\varphi$ with $\mathbb{E}_{\theta_{0},\mu}\left[\varphi\right]=\alpha$
for all $\mu\in\mathcal{H}_{\mu}$,
$
\int \mathbb{E}_{\theta,\mu}\left[\varphi^{*}\left(\theta_{0}\right) -\varphi\right]d\pi\left(\theta,\mu\right)\ge0.
$

\end{theorem}

Theorem \ref{thm: testing} states that optimal similar tests reject when an integrated likelihood ratio exceeds a conditional critical value.
Implementation is straightforward so long as one can efficiently evaluate the integrated likelihood ratio statistic $T=t(g)$ for a given realization of $g(\cdot)$.  Specifically, to simulate the conditional critical value $c_{\alpha}(h)$, we draw $\xi^*\sim N(0,{\Sigma}(\theta_0,\theta_0))$ and take $T^*=t(g^*)$,
where
$
g^*(\cdot)=h(\cdot)+{\Sigma}(\cdot,\theta_0){\Sigma}(\theta_0,\theta_0)^{-1}\xi^*.
$
The conditional critical value $c_{\alpha}(h)$ is the $1-\alpha$ quantile of $T^*$, and the test rejects if $T$ exceeds $c_{\alpha}(h)$.

 Note that while the test $\varphi^{*}\left(\theta_{0}\right)$
depends on the weight function $\pi$, it controls the rejection probability
for all parameter values consistent with the null. Hence, if we form a confidence set by collecting the non-rejected values,
$$CS=\left\{ \theta\in\Theta:\varphi^{*}\left(\theta\right)=0\right\} =\left\{\theta:\ell^{*}\left(\theta\right)\ge\frac{\int\ell^{*}\left(\tilde\theta\right)\pi\left(\tilde\theta\right)d\tilde\theta}{c_{\alpha}(h)}\right\},$$
this set has frequentist coverage $1-\alpha$ for $\theta^*$ no matter the choice of $\pi$. Thus, while we use priors to direct power, this confidence set is valid in the usual frequentist sense.  Comparing $CS$ to the level $1-\alpha$ highest quasi-posterior density credible region,
$$CR=\left\{ \theta\in\Theta:\ell^{*}\left(\theta\right)\ge\frac{\int\ell^{*}\left(\tilde\theta\right)\pi\left(\tilde\theta\right)d\tilde\theta}{\kappa(g)\pi(\theta)}\right\},$$
for $\kappa(g)^{-1}$ the $\alpha$-quantile of the posterior density $\frac{\ell^*(\theta)\pi(\theta)}{\int\ell^{*}\left(\tilde\theta\right)\pi\left(\tilde\theta\right)d\theta}$, we see that the two sets take a similar form, thresholding  $\ell^*(\theta),$ where the confidence set chooses the threshold to ensure correct coverage.  In this sense, the confidence set is a natural complement to the credible region, modifying the threshold to ensure correct coverage.

\section{Feasible Procedures}\label{sect - implementation}
The limit experiment studied in the previous sections treats $\Theta_0$ as known. In practice, however, the structure of weak identification, and thus the set $\Theta_0$, is often unknown. Moreover, we don't observe the limiting process $g(\cdot),$ and don't know $\Sigma.$
Feasible quasi-Bayes procedures replace them by the normalized sample moments $g_n(\cdot)=\frac{1}{\sqrt{n}}\sum_{i=1}^n\phi(X_i,\cdot)$ and estimated covariance $\widehat\Sigma_n$.  The researcher specifies a prior over the whole parameter space $\Theta$, and for $Q_n(\theta)=g_{n}\left(\theta\right)'\widehat{\Sigma}_n^{-1} \left(\theta,\theta\right)g_{n}\left(\theta\right)$ uses the decision rule
\begin{align}\label{eq: Bayes decision rule}
s_n(g_n)=\arg\min_{a\in\mathcal{A}}\frac{\int_{\Theta}L(a,\theta)\pi(\theta) \exp\left\{-\frac{1}{2}Q_n\left(\theta\right)\right\}d\theta}{\int_{\Theta} \pi(\theta)\exp\left\{-\frac{1}{2}Q_{n}\left(\theta\right)\right\}d\theta}.
\end{align}

This section shows that feasible decision rules (\ref{eq: Bayes decision rule}) are asymptotically equivalent to infeasible rules based on knowledge of $\Theta_0$.
In particular, the feasible quasi-posterior
$\pi(\widetilde{\Theta} |g_n)=\frac{\int_{\widetilde{\Theta}}\pi(\theta) \exp\left\{-\frac{1}{2}Q_{n}\left(\theta\right)\right\}d\theta}{\int_{\Theta} \pi(\theta)\exp\left\{-\frac{1}{2}Q_{n}\left(\theta\right)\right\}d\theta}$
concentrates on neighborhoods of $\Theta_0$.\footnote{Liao and Jiang (2010) establish a similar consistency result for the case where the weighting matrix does not vary with $\theta$.} Moreover, there exist infeasible decision rules that are asymptotically equivalent to (\ref{eq: Bayes decision rule}) for a large class of loss functions. These rules correspond to a prior $\pi^0$ supported on $\Theta_0$ and a transformation of the moments.  Specifically, some of the original moments are used to estimate $\Theta_0$, while the remainder are used to form the posterior on $\Theta_0$. We focus on unconstrained decision rules for brevity, but a similar analysis applies for tests, and sufficient conditions for uniform asymptotic validity of tests are provided in Andrews and Mikusheva (2016).

To derive these results, we first assume that the moments, appropriately centered and scaled, converge to a Gaussian process with consistently estimable variance.
\begin{ass}\label{ass:asymptotics}
The distribution $P_0$ is such that $\Phi\left(\theta\right)=\mathbb{E}_{P_0}\left[\phi\left(X_{i};\theta\right)\right]$ and $\Sigma(\cdot,\cdot)$
are continuous, and the determinant of $\Sigma(\theta)=\Sigma(\theta,\theta)$ is nonzero. Further, under $P_0$
\[
G_n(\cdot)=g_{n}\left(\cdot\right)-\sqrt{n}\Phi\left(\cdot\right)\Rightarrow G(\cdot)\sim\mathcal{GP}\left(0,\Sigma\right),
\]
and the covariance estimator $\widehat{\Sigma}_n$
is uniformly consistent,
$
\sup_{\theta\in\Theta}\left\Vert \widehat{\Sigma}_n\left(\theta\right)-\Sigma\left(\theta\right)\right\Vert \to_{p}0.
$
\end{ass}

We next assume that, on a neighborhood of $\Theta_0$, the model can be reparameterized in terms of strongly identified parameter $\gamma$ and weakly or partially identified parameter $\beta$.

\begin{ass}\label{ass:reparameterization} There exists a  continuously differentiable function $\vartheta(\beta,\gamma):\Xi\to\Theta^*\subseteq\Theta$ where $\Theta_0\subseteq\Theta^*$,  $\Xi=\left\{ \left(\beta,\gamma\right):\beta\in B,\gamma\in\Gamma\left(\beta\right)\subseteq\mathbb{R}^{p_\gamma}\right\} $ is compact, $\vartheta\left(\beta,\gamma\right)\in\Theta_{0}$
if and only if $\gamma=0$, and $0$ lies in the interior of $\Gamma\left(\beta\right)$ for all $\beta\in B$.
There exist a (positive) measure $\pi\left(\beta,\gamma\right)$ on $\Xi$
such that $\pi\left(\theta\right)$ on $\Theta^*$ is the pushforward of $\pi\left(\beta,\gamma\right)$
under $\vartheta$.  The conditional prior on $\gamma$
given $\beta$ has a uniformly  bounded density $\pi_\gamma\left(\gamma|\beta\right)$ that is uniformly 
continuous and  positive at $\gamma=0$, and $\int_B d\pi(\beta)>0$.
\end{ass}
Finally, we shorthand $\Phi(\beta,\gamma)=\Phi(\vartheta(\beta,\gamma))$ and $\Phi(\beta)=\Phi(\beta,0)$ for all functions.
\begin{ass}\label{ass: Expansion} The function $\Phi(\beta,\gamma)$ is uniformly (over $\beta\in B$) differentiable in $\gamma$ at $\gamma=0$. Further, for $\nabla\left(\beta\right)=\frac{\partial}{\partial\gamma}\Phi\left(\beta\right)$,  $J\left(\beta\right)=\frac{1}{2}\nabla\left(\beta\right)' \Sigma\left(\beta\right)^{-1}\nabla\left(\beta\right)$ is everywhere positive definite.
\end{ass}

Assumption \ref{ass:asymptotics} is standard for asymptotic analysis. Assumption \ref{ass:reparameterization} imposes that on a neighborhood of $\Theta_0$ there exists some (unknown to the researcher) re-parameterization  of the model in terms of $\beta$ and $\gamma$, where $\beta$ indexes the weakly or partially identified parameter, while $\gamma$ can be called strongly identified.  The set $\Theta_0$ corresponds to $\gamma=0,$ and is parameterized by $\beta\in B$.
Han and  McCloskey (2019) provide sufficient conditions for such a reparameterization to exist.
The mapping from $(\beta,\gamma)$ to $\theta$ can be many-to-one, and we impose very little structure on the set $B$, which may, for example, be a collection of points or intervals. We also note that $\pi(\beta,\gamma)$ need not integrate to one, since $\Theta^*$ may be a strict subset of $\Theta$. Assumption \ref{ass: Expansion} requires that $\gamma$ be strongly identified, in the sense that the Jacobian of the moments with respect to $\gamma$ has full rank at $\gamma=0$.

\begin{theorem}\label{thm: consistency} Suppose Assumptions \ref{ass:asymptotics}, \ref{ass:reparameterization}, and \ref{ass: Expansion} hold.
If the prior $\pi(\theta)$ has bounded density  on $\Theta$, then for any sequence $c_n\to\infty$, under sequences $ P_{n,f}$ local to $P_0$ in the sense of (\ref{eq: Path restriction}) we have
\begin{align}\label{eq: posterior consistency}
\pi\left(\left\{\theta\in \Theta: \Phi(\theta)\Sigma(\theta)^{-1}\Phi(\theta)\geq \frac{c_n}{n}\right\}|g_n\right)=o_{p}(1).
\end{align}
Moreover, for any bounded function $c(\theta)$ uniformly continuous at $\Theta_0$,
\begin{align}\label{eq: convergence of posterior integrals}
\int_\Theta c(\theta)d\pi(\theta|g_n)-\frac{\int_{B}c(\vartheta(\beta)) \exp\left\{-\frac{1}{2}Q_n^\beta(\beta)\right\}d\pi^0(\beta)}{\int_{B} \exp\left\{-\frac{1}{2}Q_n^\beta(\beta)\right\}d\pi^0(\beta)}=o_{p}(1),
\end{align}
where $d\pi^0(\beta)=\pi_\gamma(0|\beta)|J(\beta)|^{-\frac{1}{2}}d\pi(\beta)$, $Q_n^\beta(\beta)=g_{n}\left(\beta\right)'M \left(\beta\right)g_{n}\left(\beta\right)$, and
\[
M\left(\beta\right)=\Sigma\left(\beta\right)^{-1}-\Sigma\left(\beta\right)^{-1} \nabla\left(\beta\right )J(\beta)^{-1}\nabla\left(\beta\right)'\Sigma\left(\beta\right)^{-1}.
\]
\end{theorem}

Theorem \ref{thm: consistency} is a version of the Bernstein-von Mises theorem for weakly and partially identified quasi-Bayesian settings. The GMM objective function $Q_n(\theta)$ is bounded on $\Theta_0$ but diverges away from $\Theta_0$. As (\ref{eq: posterior consistency}) highlights, this forces the quasi-posterior to concentrate on infinitesimal neighborhoods of $\Theta_0$, corresponding to consistent estimation of the strongly identified parameter $\gamma$. The rank $k-p_\gamma$ matrix $M(\beta)$ then selects the linear combination of moments orthogonal to those used to estimate $\gamma$, and this combination is used to form the posterior on $\beta$. Unlike in the classical Bernstein-von Mises theorem, the prior on $\Theta_0$ (i.e. on $B$) matters asymptotically,  and is adjusted based on the precision of the estimate for $\gamma$ as measured by $J(\beta)$.\footnote{See Kleibergen and Mavroeidis (2014) for related discussion.} Overall, we obtain that feasible quasi-Bayes posteriors are asymptotically equivalent to infeasible posteriors based on a transformation of the prior and moment conditions.  This likewise implies asymptotic equivalence of feasible and infeasible decision rules.

\begin{corollary}\label{cor: decisions} Let the assumptions of Theorem \ref{thm: consistency} hold.
Assume that the loss function $L(a,\theta)$ is Lipschitz in $a$ and continuous in $\theta$ over $\Theta^*$, and that $\mathcal{A}$ is compact. Assume further that for almost all realization of process $G(\beta)\sim \mathcal{GP} (0,\Sigma)$, the process $L(a)=\int_{B} L(a,\vartheta(\beta))\exp\left\{-\frac{1}{2}G\left(\beta\right)'M \left(\beta\right)G\left(\beta\right)\right\}d\pi^0(\beta)$ has a unique minimizer over $\mathcal{A}$. Then
\begin{align*}
s_n^0(g_n)=\arg\min_{a\in\mathcal{A}}\frac{\int_{B} L(a,\vartheta(\beta))\exp\left\{-\frac{1}{2}Q_n^\beta(\beta)\right\}d\pi^0(\beta)}{\int_{B} \exp\left\{-\frac{1}{2}Q_n^\beta(\beta)\right\}d\pi^0(\beta)}\to^ps_n(g_n).
\end{align*}
\end{corollary}
Uniqueness of the minimizer $L(a)$ is guaranteed to hold if the loss function
is convex in $a$. Sufficient conditions for uniqueness in non-convex
cases are discussed in Cox (2020).

Overall, we obtain that feasible quasi-Bayes decision rules, computed without knowledge of $\Theta_0,$ converge to infeasible quasi-Bayes rules based on knowledge of $\Theta_0$ and a transformation of the moments and prior.  These rules are, in turn, the limit of sequences of proper-prior Bayes decision rules in the limit problem by our previous results.

\paragraph{Example: Quantile IV (continued)}
As discussed above, in this example $\Theta_0=\left\{\theta=(\alpha,\beta):\alpha=q_{P_0,\tau}(Y-W'\beta)\right\}$.  This set is one-dimensional, and can be parameterized by $\beta$. We may then define the strongly identified parameter as $\gamma=\alpha-q_{P_0,\tau}(Y-W'\beta),$ so $\vartheta(\beta,\gamma)=(\gamma+q_{P_0,\tau}(Y-W'\beta),\beta),$ and $\vartheta(\beta,\gamma)\in\Theta_0$ if and only if $\gamma=0,$ as required for Assumption \ref{ass:reparameterization}. The remainder of Assumptions \ref{ass:asymptotics}-\ref{ass: Expansion} hold under regularity conditions on $P_0$ and $\pi(\theta)$. Hence, Theorem \ref{thm: consistency} and Corollary \ref{cor: decisions} imply that feasible quasi-Bayes decision rules (\ref{eq: Bayes decision rule}), formed without knowledge of $\Theta_0$, are asymptotically equivalent to rules formed with knowledge of $P_0$ and a reduced set of moments. $\Box$

\section{Empirical Illustration}\label{section: empirical}
\begin{figure}
\includegraphics[scale=0.47]{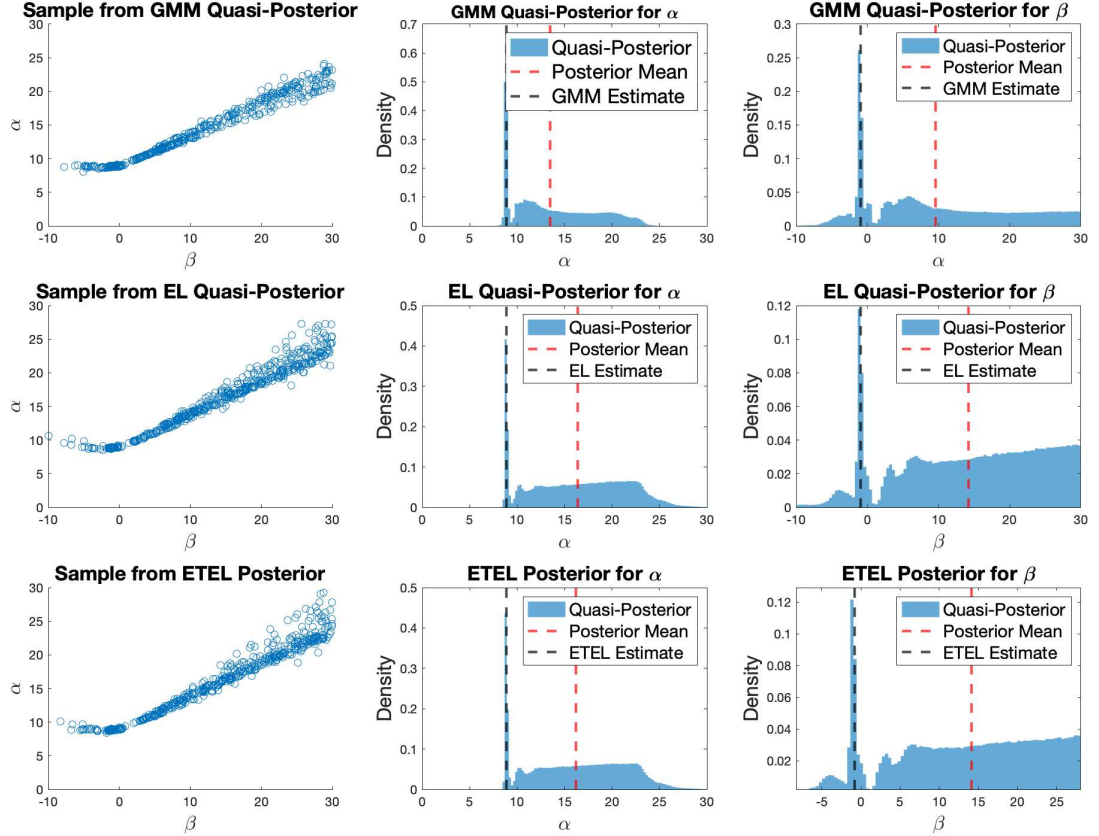}
\caption{Joint and marginal quasi-posteriors for quantile IV, $\tau=0.75,$ based on Graddy (1995) data. The results in the first row are based on the continuously updated GMM objective following Chernozhukov and Hong (2003), while the second row is empirical likelihood quasi-posterior proposed by Lazar (2003), and the third row is the exponentially tilted empirical likelihood posterior proposed by Schennach (2005)\label{fig: GEL Fish Results}}
\end{figure}

This section returns to quantile IV in the Graddy (1995) data, and reports the results from a quasi-Bayesian analysis.
Following Chernozhukov et al. (2009), we restrict
attention to $\alpha\in[0,30]$ and $\beta\in[-10,30].$\footnote{This choice of parameter space is discussed in the working paper version, Chernozhukov
et al. (2006).} We  use a flat prior $\pi\left(\theta^{*}\right)$ on $\theta=(\alpha,\beta)$, and calculate the quasi-Bayes posterior distribution discussed in Section \ref{sect - implementation}  via slice sampling (Neal, 2003).
The first row of Figure \ref{fig: GEL Fish Results} plots a sample from the joint posterior, along with the marginals for $\alpha$ and $\beta$.  Three features emerge clearly: first, the quasi-posterior distribution is highly non-normal, and is indeed bimodal, consistent with weak identification.  Second, the quasi-posterior mean differs considerably from the GMM estimate, again consistent with weak identification. Third, despite weak identification of $\alpha$ and $\beta$ separately, the quasi-posterior is concentrated around a lower-dimensional set, as expected given the posterior consistency for $\Theta_0$ established by Theorem \ref{thm: consistency}.

\begin{figure}
\includegraphics[scale=0.45]{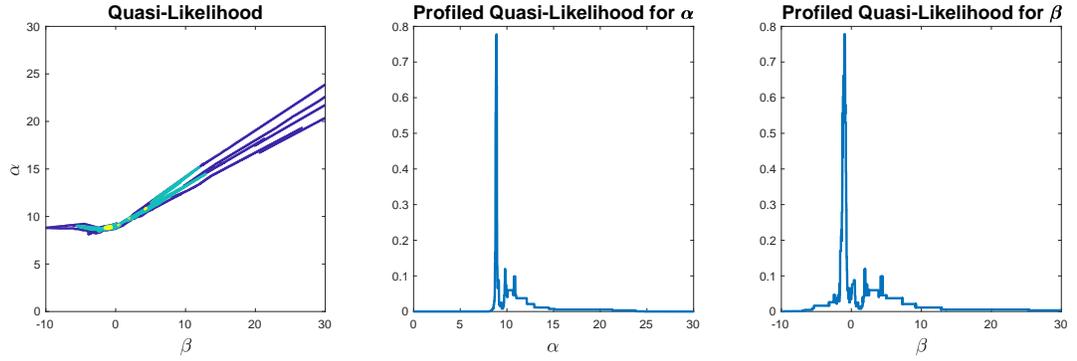}
\caption{Quasi-likelihood $\exp(-\frac{1}{2}Q_n(\theta))$ for quantile IV, $\tau=0.75,$ based on Graddy (1995) data.  Note that this is a transformation of the GMM objective plotted in Figure \ref{fig: GMM objective plot}.  \label{fig: objective plot}}
\end{figure}

The posterior bimodality observed in Figure \ref{fig: GEL Fish Results} stems from a combination of two forces.  First, if we consider the profiled quasi-likelihood  for $\alpha$ and $\beta$ separately, as shown in the second and third panels of  Figure \ref{fig: objective plot}, we see that it is multimodal, albeit with a pronounced global maximum at the GMM estimate.  As is clear from inspection, however, both profiled quasi-likelihoods are less right-skewed then their respective quasi-posteriors.  The additional skewness of the quasi-posterior reflects that the quasi-likelihood becomes less steeply curved in the strongly identified direction as $\alpha$ and $\beta$ increase, and so remains high over a wider region, as shown in the first panel of Figure \ref{fig: objective plot}.  When we integrate against the prior, this increases the posterior mass assigned to large values of $\alpha$ and $\beta$.  This effect is captured in the $|J(\beta)|^{-\frac{1}{2}}$ term in Theorem \ref{thm: consistency}, which shows that quasi-posterior assigns more mass to regions where the strongly-identified parameter is less precisely estimated.

While we have focused on quasi-Bayes procedures (\ref{eq: Bayes decision rule}) based on the continuously updating GMM objective function, as noted in Section  \ref{sec: quasi-posterior rules} quasi-Bayes procedures based on other GEL-type objectives are equivalent in the limit experiment.  The second and third rows of Figure  \ref{fig: GEL Fish Results} plot the the empirical likelihood quasi-posterior proposed by Lazar (2003) and the exponentially tilted empirical likelihood posterior proposed by Schennach (2005), along with the corresponding estimators. In both cases the results are qualitatively  similar to those obtained based on continuously updating GMM.

\begin{figure}
\includegraphics[scale=0.42]{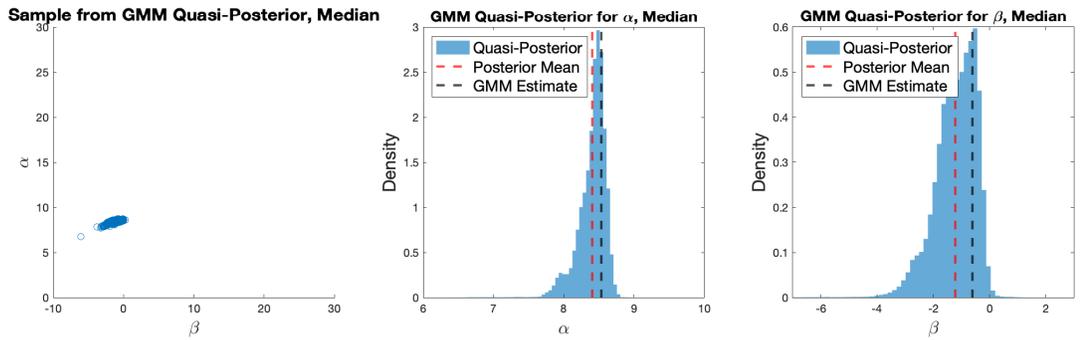}
\caption{Joint and marginal quasi-posteriors for quantile IV for median, $\tau=0.5,$ based on Graddy (1995) data\label{fig: Median Fish Results}}
\end{figure}

Finally, to illustrate the impact of varying identification strength on the quasi-posterior, we consider inference on the median $\tau=0.5$, which might be expected to be more strongly identified than $\tau=0.75.$  Consistent with this intuition, the quasi-posterior for the median (plotted in Figure \ref{fig: Median Fish Results}) is much more concentrated, and the GMM estimate is closer to the quasi-posterior mean.  Interestingly, however, even in this case the quasi-posterior is clearly non-normal, suggesting that strong-identification asymptotic approximations may not be fully reliable.  Consistent with this possibility, in simulations calibrated to this application (and reported in the  Appendix) we find that the distribution of the GMM estimator is clearly non-normal, and that the weak-identification asymptotic approximation again better matches the finite-sample distribution.

While one could use the quasi-posteriors discussed above to form credible sets, as discussed in Section \ref{sec: tests} these sets will not in general have correct frequentist coverage.
 To ensure coverage, one can instead compute identification-robust confidence sets.  In this application the resulting confidence set, shown in Appendix D, closely resembles the 95\% highest posterior density set.

\section{References}

\noindent Adler, R.J., and J.E. Taylor (2007), \emph{Random Fields and Geometry}, New York : Springer

\noindent Andrews, D.W.K., M. Moreira, and J. Stock (2006), ``Optimal Two-Sided Invariant Similar Tests of Instrumental Variables Regression,'' \emph{Econometrica}, 74(3), 715-752

\noindent Andrews, I. and A. Mikusheva (2016), ``Conditional Inference with a Functional Nuisance Parameter,''  \emph{Econometrica}, 84(4), 1571-1612

\noindent Berlinet, A., and C. Thomas-Agnan (2004) \emph{Reproducing Kernel Hilbert Spaces in Probability and Statistics}, Kluwer Academic Publishers

\noindent Bornn, L., N. Shephard, and R. Solgi (2019) ``Moment Conditions and Bayesian Nonparametrics,'' \emph{Journal of the Royal Statistical Society, Series B}, 81: 5-43

\noindent Brown, L.D.  (1986) \emph{Fundamentals of Statistical Exponential Families with Applications in Statistical Decision Theory}, Hayward, CA: Institute of Mathematical Statistics

\noindent  Chamberlain, G. and G. Imbens (2003): ``Nonparametric Applications of Bayesian Inference,'' \emph{Journal of Business and Economic Statistics},  21, 12–18

\noindent Chen, X., T. Christensen, and E. Tamer (2018). ``Monte Carlo Confidence Sets for Identified Sets,''
\emph{Econometrica}, 86(6), 1965-2018

\noindent  Chernozhukov, V. and C. Hansen (2005): ``An IV Model of Quantile
Treatment Effects,'' \emph{Econometrica}, 73(1), 245-261

\noindent Chernozhukov, V., C. Hansen, and M. Jansson (2006), ``Finite Sample
Inference for Quantile Regression Models,'' Unpublished Manuscript

\noindent Chernozhukov, V., C. Hansen, and M. Jansson (2009), ``Finite Sample
Inference for Quantile Regression Models,'' \emph{Journal of Econometrics},
152(2), 93-103

\noindent  Chernozhukov, V. and H. Hong (2003): ``An MCMC Approach to Classical Estimation,'' \emph{Journal of Econometrics}, 115(2), 293-346

\noindent Cox, G. (2020), ``Almost Sure Uniqueness of a Global Minimum Without
Convexity,'' \emph{Annals of Statistics}, 48(1), 584-606

\noindent  Florens, J.P. and A. Simoni  (2019): ``Gaussian Processes and Bayesian Moment Estimation,'' \emph{Journal of Business and Economic Statistics}, Forthcoming.

\noindent Gallant, R. (2016), ``Reflections on the Probability Space Induced by Moment Conditions with Implications for Bayesian Inference,'' \emph{Journal of Financial Econometrics}, 14(2), 284–294

\noindent Graddy, K. (1995), ``Testing for Imperfect Competition at the Fulton
Fish Market,'' \emph{Rand Journal of Economics,} 26(1), 75-92

\noindent Guggenberger, P. and R. J. Smith (2005), ``Generalized Empirical Likelihood Estimators and Tests Under Partial, Weak, and Strong Identification,'' \emph{Econometric Theory,} 21(4), 667-709

\noindent Han, S. and A. McCloskey (2019), ``Estimation and Inference with
a (Nearly) Singular Jacobian,'' \emph{Quantitative Economics}, 10(3),
1019-1068

\noindent Hirano, K. and J. Porter (2009), ``Asymptotics for Statistical Treatment Rules,''  \emph{Econometrica}, 77(5), 1683-1701

\noindent  Kaji, T.  (2020): ``Theory of Weak Identification in Semiparametric Models,'' \emph{Econometrica}, Forthcoming

\noindent  Kim, J.Y.  (2002): ``Limited Information Likelihood and Bayesian Analysis,'' \emph{Journal of Econometrics}, 107, 175-193

\noindent  Kitamura, Y. and T. Otsu (2011),  ``Bayesian Analysis of Moment Restriction Models Using Nonparametric Priors,''  Unpublished Manuscript

\noindent Kleibergen, F. and S. Mavroeidis (2014): ``Identification Issues in Limited-Information Bayesian Analysis of Structural Macroeconomic Models,'' \emph{Journal of Applied Econometrics}, 29(7), 1183-1209

\noindent  Lazar, N. A.  (2003): ``Bayesian Empirical Likelihood,'' \emph{Biometrika}, 90(2), 319-326

\noindent Le Cam, L. (1986), \emph{Asymptotic Theory of Statistical Inference,} John Wiley \& Sons

\noindent Lehman, E.L. and G. Casella (1998),\emph{ Theory of Point Estimation,} Springer Texts in Statistics

\noindent Liao, Y. and W. Jiang (2010), ``Bayesian Analysis in Moment Inequality Models,'' \emph{Annals of Statistics}, 38(1), 275-316

\noindent Moon, H.R. and F. Schorfheide (2012), Bayesian and Frequentist Inference in Partially Identified Models. \emph{Econometrica,} 80(2), 755-782.

\noindent Moreira, H. and M. Moreira (2019), ``Optimal Two-Sided Tests for Instrumental Variables Regression with Heteroskedastic and Autocorrelated Errors,'' \emph{Journal of Econometrics}, 213(2), 398-433

\noindent Moreira, M. and G. Ridder (2019), ``Optimal Invariant Tests in an Instrumental Variables Regression With Heteroskedastic and Autocorrelated Errors,'' Unpublished Manuscript

\noindent Mueller, U.K. (2011), ``Efficient Tests Under a Weak Convergence Assumption,'' \emph{Econometrica}, 79(2), 395-435

\noindent Neal, R.M. (2003), ``Slice Sampling,'' \emph{Annals of Statistics}, 31(3), 705-767

\noindent  Neveu J . (1968) \emph{Processus AIeatoires Gaussiens.} Seminaire Math. Sup., Les presses de
I'Universite de Montreal

\noindent Parzen, E. (1962), ``Extraction and Detection Problems and Reproducing Kernel Hilbert Spaces,'' \emph{J. SIAM Control}, Ser. A., 1(1) , 35--62

\noindent Schennach, S. M. (2005), ``Bayesian Exponentially Tilted Empirical Likelihood,'' \emph{Biometrika} 92, 31–46

\noindent  Shin,  M. (2015), ``Bayesian GMM,'' Dissertation Chapter, University of Pennsylvania


\noindent  Strasser, H.  (1985), \emph{Mathematical  Theory  of  Statistics:  Statistical  Experiments and  Asymptotic  Decision  Theory}  (De  Gruyter  Studies  in  Mathematics  7). Berlin: de Gruyter.

\noindent  Van der Vaart, A.W. (1991)  ``An Asymptotic Representation Theorem,'' \emph{International Statistical Review}, Vol. 59, No. 1, pp. 97-121

\noindent  Van der Vaart, A.W. (1998)   \emph{Asymptotic Statistics}, Cambridge University Press

\noindent  Van der Vaart, A.W. and J.A. Wellner (1996),  \emph{Weak Convergence and Empirical Processes}, Springer

\noindent  Van der Vaart, A.W. and H. Van Zanten (2008), ``Reproducing Kernel Hilbert Spaces of Gaussian Processes,'' in \emph{Pushing the Limits of Contemporary Statistics: Contributions in Honor of Jayanta K. Ghosh},  Bertrand Clarke and Subhashis Ghosal, eds., (Beachwood, Ohio, USA: Institute of Mathematical Statistics, 2008), 200-222

 \subsection*{Appendix A: Proofs}

\begin{definition}  The Reproducing Kernel Hilbert Space
(RKHS) $\mathcal{H}$ associated with $\Sigma$  is a space of functions from $\Theta_0$ to $\mathbb{R}^{k}$, which is obtained as the completion
of the space spanned by functions  of the form $\sum_{j=1}^{s}\Sigma(\cdot,\theta_{j})b_{j}$ for any finite set of constant vectors $\{b_{j}\}\subset\mathbb{R}^{k}$ and parameters $\{\theta_{j}\}\subset\Theta_0$,
with respect to the scalar product $\langle\cdot,\cdot\rangle_{\mathcal{H}}$ defined on the basis set as
$$
\langle\sum_{j=1}^{s}\Sigma(\cdot,\theta_{j})b_{j},\sum_{l=1}^{\tilde{s}}\Sigma(\cdot,\tilde{\theta}_{l})c_{l}\rangle_{\mathcal{H}} =\sum_{j=1}^{s}\sum_{l=1}^{\tilde{s}}b_{j}'\Sigma(\theta_{j},\tilde{\theta}_{l})c_{l}.
$$
\end{definition}

\begin{lemma}\label{lem: on correspondence of rkhs}
The image of $T(P_0)$ under  transformation (\ref{eq:isomorphism}) is $\mathcal{H}$.
\end{lemma}
\paragraph{Proof of Lemma \ref{lem: on correspondence of rkhs}} A score $f(X)=\sum_{i=1}^s\phi(X,\theta_i)'a_i$ corresponds to the  function
$$
m(\cdot)=\mathbb{E}_{P_0}[f(X)\phi(X,\cdot)]=\sum_{i=1}^s\mathbb{E}_{P_0}[\phi(X,\cdot)\phi(X,\theta_i)'a_i]=\sum_{i=1}^s\Sigma(\cdot,\theta_i)a_i.
$$
For two scores $f_1(X)=\sum_{i=1}^s\phi(X,\theta_i)'a_i$ and $f_2(X)=\sum_{j=1}^{s^*}\phi(X,\theta^*_j)'b_j$ and corresponding mean functions $m_1(\cdot)=\sum_{i=1}^s\Sigma(\cdot,\theta_i)a_i$ and $m_2(\cdot)=\sum_{j=1}^{s^*}\Sigma(\cdot,\theta^*_j)b_j$ we have
\begin{align*}
\mathbb{E}_{P_0}\left[f_1(X)f_2(X)\right]=\sum_{i=1}^s\sum_{j=1}^{s^*} a_i'\mathbb{E}_{P_0}\left[\phi(X,\theta_i)\phi(X,\theta^*_j)' \right]b_j=\sum_{i=1}^s\sum_{j=1}^{s^*} a_i'\Sigma(\theta_i,\theta^*_j)b_j=\langle m_1, m_2\rangle_{\mathcal{H}}.
\end{align*}
This implies that there is an isomorphism between $\mathcal{H}$ and  the completion of the
space spanned by scores of the form $f(X)=\sum_{j=1}^{s}\phi(X,\theta_{j})'b_{j}$ in
$L_{2}({P_0})$.

It remains to show that for any $f^\bot\in T({P_0})$  that is orthogonal to the completion of the
space spanned by scores of the form $f(X)=\sum_{j=1}^{s}\phi(X,\theta_{j})'b_{j}$ in
$L_{2}({P_0})$ we have $\mathbb{E}_{P_0}[f^{\bot}(X)\phi(X,\cdot)]\equiv 0\in \mathbb{R}^k$. Indeed, $f^\bot$ is orthogonal to  $a'\phi(X,\theta)$ for any vector $a\in \mathbb{R}^k$ and any $\theta\in \Theta_0$. Thus $\mathbb{E}_{P_0}[f^{\bot}(X)a'\phi(X,\theta)]=a'm(\theta)=0$. This also shows that $\mathcal{H}^*$ coincides with the completion of the
space spanned by scores of the form $f(X)=\sum_{j=1}^{s}\phi(X,\theta_{j})'b_{j}$ in
$L_{2}({P_0})$.

\begin{definition} Let $\{\varphi_{j}\}$ be
a complete orthonormal basis in $T(P_0)$. Define an experiment $\mathcal{E}_{\infty}^{*}$ as the one of observing the  (infinite) sequence of independent random variables
$
W_{j}\sim N\left(\mathbb{E}_{P_0}[f(X)\varphi_{j}(X)],1\right),
$ with parameter space $\left\{(\theta^*,f):\theta^*\in\Theta_0,f\in T_{\theta^*}(P_0)\right\}$.
\end{definition}

\begin{lemma}\label{lem: representation theorem}(Theorem 3.1 in Van der Vaart, 1991)
Consider a sequence of statistics $S_{n}$ which has a limit distribution
under $\mathcal{E}_{n}^{*},$ in the sense that under any $P_{n,f}$ for $f\in T(P_0),$
$
S_{n}\left(X_{1},...,X_{n}\right)\Rightarrow S_{f}
$
as $n\to\infty.$
Assume there exists a complete separable set \textbf{$\mathbb{S}_{0}$}
such that $S_{f}(\mathbb{S}_{0})=1$ for all $f\in T(P_0)$. Then in the experiment $\mathcal{E}_{\infty}^*$ there
exists a (possibly randomized) statistic
$
S^{*}=s^*\left(\left\{ W_{j}\right\} ,U\right)
$
for a random variable $U\sim U\left[0,1\right]$  independent of $W_{j}$
such that
$
S^{*}\sim S_{f}$ under $f$ for all $f\in T(P_0).$

\end{lemma}

\paragraph{Proof of Theorem \ref{thm: GP representation theorem}}
Define the orthonormal basis $\{\varphi_j(X)\}$ of $T(P_0)$ to consist of the union of an orthonormal basis $\{\varphi_j^*(X)\}$ of $\mathcal{H}^*$ and an orthonormal basis $\{\varphi_j^{\bot}(X)\}$ of $\mathcal{H}^\bot$. The limit experiment $\mathcal{E}^*_\infty$ corresponds to observing the union of two sets of mutually independent random variables:
$$
W_j^*\sim N\left(\mathbb{E}_{P_0}[f(X)\varphi_{j}^*(X)] ,1\right) \mbox { and }  W_j^\bot \sim N\left(\mathbb{E}_{P_0}[ f(X)\varphi_{j}^\bot(X)] ,1\right).
$$
Due to Lemma \ref{lem: on correspondence of rkhs} we have $\mathbb{E}_{P_0}[ f(X)\varphi_{j}^*(X)]=\langle m, \varphi_{j}^*\rangle_\mathcal{H}$. The experiment of observing only
$
W_j^*\sim N\left(\langle m, \varphi_{j}^*\rangle_\mathcal{H} ,1\right)
$
is equivalent to the Gaussian Process experiment $\mathcal{E}_{GP}^*$ (see e.g. Theorem 4.3 of van der Vaart and van Zanten, 2008).

By independence $dP_f(W^*,W^\perp)=dP_{f^*}(W^*)\times dP_{f^\perp}(W^\perp)$.
The loss function depends only on $\theta^*$, and the parameter space for $(\theta^*,f^*,f^{\bot})$ is the Cartesian product $\{\theta^*\in \Theta_0,f^*\in \mathcal{H}^*_{\theta^*}\}\times\{f^\bot\in \mathcal{H}^\bot\}$.  The risk of a decision rule $\delta$ is
$$
\widetilde{R}(\theta^*,f)=\widetilde{R}(\theta^*,f^*,f^\perp)=\mathbb{E}_f\left[L(\delta(W^*,W^\perp),\theta^*)\right].
$$
We claim that for any fixed value $f^\perp$ there exists a decision rule in the experiment $\mathcal{E}_{GP}^*$ with risk
$
R(\theta^*,m)=\widetilde{R}(\theta^*,f^*,f^\perp) $ for all $(\theta^*,f^*)\in\{\theta^*\in \Theta_0,f^*\in \mathcal{H}^*_{\theta^*}\}$, where $m$ corresponds to $f^*$.  Indeed, since experiment $\mathcal{E}_{GP}^*$  is equivalent to observing only the $W^*_j$ variables, it is enough for each realization $W^*=w$ to draw a random variable $(W^\perp)$ from distribution $dP_{f^\perp}$ (which is  fixed) and produce a randomized  decision as
$
\widetilde\delta(w)=\delta(w,W^\perp).
$
The last line of the result then follows from the portmanteau lemma.

\paragraph{Proof of Theorem \ref{Thm: Brown admissibility}}
The distribution of $g$ for any $m\in\mathcal{H}$ is dominated by the distribution under $m=0$.
Moreover, the form of the likelihood ratio for Gaussian processes (see e.g. Theorem 54 in Berlinet and Thomas-Agnan) implies that condition  (1) in Section 4A.1 of Brown (1986) holds.
Our assumptions likewise imply condition (2) of Brown (1986).  This result is thus immediate from Theorem 4A.12 of Brown (1986).

\paragraph{Proof of Lemma \ref{lem: reparametrization}} Denote by $G(\cdot)=g(\cdot)-m(\cdot)$ the mean-zero Gaussian process with covariance function $\Sigma$.
The regression of the process $G$ on the anchor $A(G)$ defines a Pettis integral
\begin{align}\label{eq: def psi}
\psi(\cdot)=[\psi_1(\cdot),...,\psi_k(\cdot)]=\mathbb{E}\left[G(\cdot)A(G)^\prime\right] \mathbb{E}\left[A(G)A(G)^\prime\right]^{-1}\in \mathcal{H}^k,
\end{align}
where each column is a function in $\mathcal{H}$ (see Van der Vaart and Van Zanten, 2008, for discussion) and depends only on $\Sigma$ and $A$.
Let $\mathcal{H}_\mu$ be the linear sub-space of $\mathcal{H}$ orthogonal to $\{ \psi_1(\cdot),...,\psi_k(\cdot)\}$.
For each $m(\cdot)\in \mathcal{H}$, define $\mu(\cdot)$ to be the projection of $m$ on the linear sub-space $\mathcal{H}_\mu$.
The properties of Pettis integrals imply that $\langle\psi, m\rangle_{\mathcal{H}}=\left[A(G)A(G)^\prime\right]^{-1}A(m)$ and $\langle\psi,\psi\rangle_{\mathcal{H}}=\left[A(G)A(G)^\prime\right]^{-1}$, which yields the orthogonal decomposition
$$
m(\cdot)=\mu(\cdot)+\psi(\cdot)A(m).
$$
For any $\theta^*$ and $m\in \mathcal{H}_{\theta^*}$, $m(\theta^*)=0,$ so $A(m)=-[\psi(\theta^*)]^{-1}\mu(\theta^*)$. We can consequently re-write $m(\cdot)$ as a function of $(\theta^*,\mu)$,
$$
m(\cdot)=\mu(\cdot)-\psi(\cdot)[\psi(\theta^*)]^{-1}\mu(\theta^*).
$$
This establishes a one-to-one correspondence between $\left\{(\theta^*,m):\theta^*\in\Theta_0,m\in \mathcal{H}_{\theta^*}\right\}$ and $(\theta^*,\mu)\in \Theta_0\times \mathcal{H}_\mu$.
Moreover, $\mathcal{H}_\mu$ is the RKHS generated by the covariance function
$$\widetilde{\Sigma}(\theta_1,\theta_2)=\Sigma(\theta_1,\theta_2)-\psi(\theta_1) \mathbb{E}\left[A(G)A(G)^\prime\right]\psi(\theta_2)'.$$
Define the random vector $\xi$ and stochastic process  $h$ by
$
\xi=A(g)~~\text{and}~~h(\cdot)=g(\cdot)-\psi(\cdot)\xi.
$
By construction $\xi\sim N(\nu(\theta^*,\mu),\Sigma_\xi)$ for $\nu(\theta^*,\mu)=-[\psi(\theta^*)]^{-1}\mu(\theta^*)$ and $\Sigma_\xi=\mathbb{E}[A(G)A(G)'],$ while $h(\cdot)\sim \mathcal{GP}(\mu,\widetilde\Sigma)$. Moreover, $\xi$ and $h$ are jointly normal and uncorrelated, and therefore independent.

\paragraph{Proof of Theorem \ref{thm: on invariant prior}} According to Neveu (1968) the conditional distribution of $\xi\sim N(-[\psi(\theta^*)]^{-1}\mu(\theta^*),\Sigma_\xi)$ given the realization of
$h=\mu+GP(0,\widetilde{\Sigma})$,  assuming $\mu\sim \mathcal{GP}(0,\Omega)$, is  Gaussian and the conditional mean coincides with the best linear predictor.

Let
$
\rho(\cdot)=\mathbb{E}[\xi h(\cdot)]=-[\psi(\theta^*)]^{-1}\Omega(\theta^*,\cdot),$   and  note that $  \mathbb{E}[h(\theta_1)h(\theta_2)]= \Omega(\theta_1,\theta_2)+\widetilde{\Sigma}(\theta_1,\theta_2).
$
Denote by $\mathcal{K}$ the RKHS corresponding to the covariance function $\Omega+\widetilde{\Sigma}$, and by $L(h)$ the subspace of $L_2$ random variables obtained as the closure of linear combinations of $h$.
Define $\xi^*$ as the projection of $\xi$ on to $L(h)$. By definition it is the best linear predictor of $\xi$ given $h$, and $\mathbb{E}[\xi h(\cdot)]=\mathbb{E}[\xi^* h(\cdot)]=\rho(\cdot)$. Lemma 13 in Berlinet and Thomas-Agnan (2004) implies that $\rho(\cdot)\in \mathcal{K}$. Denote by $\Psi$ the canonical congruence between $L(h)$ and $\mathcal{K}$, defined by
$$
\Psi\left(\sum_{j}a_jh(\theta_j)\right)= \sum_{j}a_j(\Omega(\theta_j,\cdot)+\widetilde{\Sigma}(\theta_j,\cdot))\in \mathcal{K},
$$
and extended by continuity. Then  $\xi^*=\Psi^{-1}(\rho(\cdot)).$ See Section 3 of Berlinet and Thomas-Agnan (2004) for further discussion.

To prove (a) we  fix $\theta^*$, assume that condition (\ref{eq: condition on omega}) holds, and show that the best linear predictor depends on $(\xi, g(\theta^*))$ only. Condition (\ref{eq: condition on omega}) implies that
$$
\Omega(\theta^*,\cdot)+\widetilde{\Sigma}(\theta^*,\cdot)= \left(I_k+\widetilde{\Sigma}(\theta^*,\theta^*)\Omega(\theta^*,\theta^*)^{-1}\right)\Omega(\theta^*,\cdot).
$$
Thus,
$$
\rho(\cdot)=-[\psi(\theta^*)]^{-1} \left(I_k+\widetilde{\Sigma}(\theta^*,\theta^*)\Omega(\theta^*,\theta^*)^{-1}\right)^{-1} \left[\Omega(\theta^*,\cdot)+\widetilde{\Sigma}(\theta^*,\cdot)\right],
$$
and the canonical congruence has the form
$$
\xi^*=\Psi^{-1}(\rho(\cdot))=-[\psi(\theta^*)]^{-1} \left(I_k+\widetilde{\Sigma}(\theta^*,\theta^*)\Omega(\theta^*,\theta^*)^{-1}\right)^{-1}h(\theta^*),
$$
which depends on the data only through $h(\theta^*)=g(\theta^*)-\psi(\theta^*)\xi$.

Now let us assume that for each $\theta^*$ the likelihood depends only on $(\xi, g(\theta^*))$ and prove that (\ref{eq: condition on omega}) holds. Since the conditional distribution of $\xi$ given $\theta^*$ and $\xi^*$ is $N(\xi^*,\Sigma_\xi)$, $\xi^*$ must depend only on $(\xi, g(\theta^*))$ or, equivalently, on $(\xi,h(\theta^*))$.  Linearity of $\xi^*$ in $h$ then implies that
 there exists a non-random $k\times k$ matrix $B(\theta^*)$ such that
$
\Psi^{-1}(\rho(\cdot))=B(\theta^*) h(\theta^*).
$
By the definition of the canonical congruence this implies
$
\rho(\cdot)=B(\theta^*)\left[\Omega(\theta^*,\cdot)+\widetilde{\Sigma}(\theta^*,\cdot)\right].
$
Since $\rho(\cdot)=-[\psi(\theta^*)]^{-1}\Omega(\theta^*,\cdot)$, however, $[\psi(\theta^*)]^{-1}\Omega(\theta^*,\cdot)=B(\theta^*)\left[\Omega(\theta^*,\cdot)+\widetilde{\Sigma}(\theta^*,\cdot)\right]$.  Since $\psi(\theta^*)$ has full rank, both sides are invertible when evaluated at $\theta^*,$ and some rearrangement yields  (\ref{eq: condition on omega}). This finishes the proof of (a)

To prove (b) let $B=\Omega(\theta_0,\theta_0)$. Condition (\ref{eq: condition on omega}) implies that
$$
\Omega(\theta_0,\theta)=B\widetilde{\Sigma}(\theta_0,\theta_0)^{-1}\widetilde{\Sigma}(\theta_0,\theta) \mbox{   and   } \Omega(\theta,\theta)^{-1}\Omega(\theta,\theta_0)=\widetilde{\Sigma}(\theta,\theta)^{-1} \widetilde{\Sigma}(\theta,\theta_0).
$$
The transposed equations are
$$
\Omega(\theta,\theta_0)=\widetilde{\Sigma}(\theta,\theta_0)\widetilde{\Sigma}(\theta_0,\theta_0)^{-1}B \mbox{   and   } \Omega(\theta_0,\theta)\Omega(\theta,\theta)^{-1}= \widetilde{\Sigma}(\theta_0,\theta)\widetilde{\Sigma}(\theta,\theta)^{-1}.
$$
We can calculate $\Omega(\theta_0,\theta)\Omega(\theta,\theta)^{-1}\Omega(\theta,\theta_0)$ in two ways, so
$$
\widetilde{\Sigma}(\theta_0,\theta)\widetilde{\Sigma} (\theta,\theta)^{-1}\widetilde{\Sigma}(\theta,\theta_0) \widetilde{\Sigma}(\theta_0,\theta_0)^{-1}B= B\widetilde{\Sigma}(\theta_0,\theta_0)^{-1}\widetilde{\Sigma}(\theta_0,\theta) \widetilde{\Sigma}(\theta,\theta)^{-1}\widetilde{\Sigma}(\theta,\theta_0).
$$
Pre and post multiply the last equation with $\widetilde{\Sigma}(\theta_0,\theta_0)^{-1/2}$ and let
$$\widetilde{B}=\widetilde{\Sigma}(\theta_0,\theta_0)^{-1/2}B\widetilde{\Sigma}(\theta_0,\theta_0)^{-1/2}.$$ We obtain that $\widetilde{B}$ commutes with a whole family of symmetric matrices :
$
D(\theta)\widetilde{B}= \widetilde{B}D(\theta).
$
Assume $\widetilde{B}$ has $r$ distinct eigenvalues. Since $\widetilde{B}$ is symmetric, all eigenvectors corresponding distinct eigenvalues are orthogonal. Let $V_1$,..., $V_r$ be the orthogonal sub-spaces spanned by eigenvectors of the $\widetilde{B}$ corresponding to eigenvalues $\lambda_1,...,\lambda_r$, respectively.  Consider a symmetric matrix $D(\theta)\in\mathcal{D}$ that commutes with $\widetilde{B}$. Take any $v_i\in V_i$ and $v_j\in V_j$:
$$
v_i'D(\theta)\widetilde{B}v_j=\lambda_jv_i'D(\theta)v_j=v_i'\widetilde{B}D(\theta)v_j=\lambda_iv_i'D(\theta)v_j.
$$
This implies $v_i'D(\theta)v_j=0$ for any $i\neq j$. Thus $D(\theta)v_j\in V_j$, and $V_1$,..., $V_r$ are invariant subspaces  for $D(\theta)$. Thus, we proved that $V_1$,...,$V_r$ are invariant spaces for the whole family of operators $\mathcal{D}$. Under the conditions of the lemma this implies that $\widetilde{B}$ has single eigenvalue $\lambda>0$, and thus $\Omega(\cdot,\cdot)=\lambda\widetilde{\Sigma}(\cdot,\cdot)$.

\paragraph{Proof of Theorem \ref{thm: testing}}

Similarity of $\varphi^{*}\left(\theta_{0}\right)$ follows from Andrews and Mikusheva (2016).  For any $\mu\in\mathcal{H}_{\mu}$ and any test $\varphi$,
\[
\int\int_{}\mathbb{E}_{\theta,\mu}\left[\varphi\right]d\pi\left(\mu\right)d\pi\left(\theta\right) =\mathbb{E}_{\pi}\left[\varphi\right]= \mathbb{E}_{\theta_{0},\mu}\left[\frac{\int\ell^{*}\left(\theta\right)d\pi(\theta) f\left(h\right)}{\ell\left(\mu,\theta_{0};\xi\right)\ell\left(\mu;h\right)}\varphi \right].
\]
Since $\xi=g\left(\theta_{0}\right),$ $\ell\left(\mu,\theta_{0};\xi\right)$
does not depend on $\mu$, and is equal to $\ell^{*}\left(\theta_{0}\right).$
Lemma 2 of Andrews and Mikusheva (2016)  implies that
$
\tilde{\varphi}\left(\theta_{0}\right)=1\left\{ \frac{\int\ell^{*}\left(\theta\right)d\pi(\theta)f\left(h\right)}{ \ell^{*}\left(\theta_{0}\right)\ell\left(\mu;h\right)}>\tilde{c}_{\alpha}\left(h\right)\right\}
$
maximizes $\mathbb{E}_{\pi}\left[\varphi\right]$ over
the class of size-$\alpha$ similar tests, where $\tilde{c}_{\alpha}\left(h\right)$
is the $1-\alpha$ quantile of $\frac{\int\ell^{*}\left(\theta\right)d\pi(\theta)f\left(h\right)}{\ell^{*}\left(\theta_{0}\right)\ell\left(\mu;h\right)}$
conditional on $h$ under the null. The test statistic in $\tilde{\varphi}\left(\theta_{0}\right)$
differs from that in $\varphi^*\left(\theta_{0}\right)$ only through
terms depending on $h$. These can be absorbed into the critical value,
so $\tilde{\varphi}\left(\theta_{0}\right)=\varphi^*\left(\theta_{0}\right).$

\paragraph{Proof of Theorem \ref{thm: consistency}} By contiguity it is enough to prove the statement under $P_0$.  Denote $Q(\theta)=\Phi(\theta)'\Sigma(\theta)^{-1}\Phi(\theta)$. Due to Assumption \ref{ass:asymptotics}, $\widehat\Sigma_n(\theta)^{-1}$ is uniformly bounded in probability and $G_n(\cdot)\Rightarrow \mathcal{GP}(0,\Sigma)$, thus  $\max_{\theta\in\Theta}G_n(\theta)'\widehat\Sigma_n(\theta)^{-1}G_n(\theta)=O_p(1)$. Since $g_n(\theta)=\sqrt{n}\Phi(\theta)+G_n(\theta)$ we have
\begin{align*}
\frac{1}{2}Q_n(\theta)\leq nQ(\theta)(1+o_p(1)) + \max_{\theta\in\Theta}G_n(\theta)'\widehat\Sigma_n(\theta)^{-1}G_n(\theta);
\\
 \frac{1}{2}Q_n(\theta)\geq \frac{n}{2}Q(\theta)(1+o_p(1)) - \max_{\theta\in\Theta}G_n(\theta)'\widehat\Sigma_n(\theta)^{-1}G_n(\theta).
\end{align*}
Define a set
$
\Theta_{\delta,n}=\left\{\theta\in \Theta: Q(\theta)\leq \frac{\delta}{n}\right\}$ for some $\delta>0$ and $\Theta_{c_n}^c=\{\theta\in \Theta: Q(\theta)\geq \frac{c_n}{n}\}$.
\begin{align*}
\pi\left(\Theta_{c_n}^c|g_n\right)\leq
\frac{ \int_{\Theta_{c_n}^c} \pi(\theta)\exp\left\{-\frac{1}{2}Q_{n}\left(\theta\right)\right\} d\theta }{\int_{\Theta_{\delta,n}} \pi(\theta) \exp\left\{-\frac{1}{2} Q_{n}\left(\theta\right)\right\} }\leq O_p(1)\cdot \frac{\int_{\Theta_{c_n}^c}\pi(\theta)\exp\{-\frac{n}{2}Q(\theta)\}d\theta}{\int_{\Theta_{\delta,n}}\pi(\theta)d\theta}.
\end{align*}
Due to uniform differentiability of $\Phi(\beta,\gamma)$  there exist positive constants $C_1, C_2$  and small enough $\varepsilon>0$ such that for all $\theta\in\Theta^*_\varepsilon=\{\theta=\vartheta(\beta,\gamma): \|\gamma\|<\varepsilon\}$ we have $C_1\|\gamma\|^2\leq Q(\beta,\gamma)\leq C_2\|\gamma\|^2$ . For large enough $n$ we have $\Theta_{\delta,n}\subseteq\Theta^*_\varepsilon$. Thus,
$$
\int_{\Theta_{\delta,n}}\pi(\theta)d\theta= \int_B\int_{Q(\beta,\gamma)\leq \frac{\delta}{n}}\pi_\gamma(\gamma|\beta)d\gamma d\pi(\beta)\geq C\int_{\|\gamma\|^2\leq \frac{\delta}{C_2n}}d\gamma\geq Cn^{-\frac{p_\gamma}{2}}.
$$
Divide the integral over $\Theta_{c_n}^c$ into integrals over $\Theta_{c_n}^c\cap \Theta^*_\varepsilon$ and over $\Theta_{c_n}^c\cap (\Theta_\varepsilon^*)^c$, where $(\Theta_\varepsilon^*)^c= (\Theta\setminus\Theta_\varepsilon^*)$.  We have $\Theta_{c_n}^c\cap \Theta_\varepsilon^*\subseteq\{\theta=\vartheta(\beta,\gamma):C_2\|\gamma\|^2\geq\frac{c_n}{n}\}$. Denote by $\bar{Q}$ the non-zero minimum of $Q(\theta)$ over $(\Theta_\varepsilon^*)^c$. Thus,
\begin{align*}
\frac{\int_{\Theta_{c_n}^c}\pi(\theta)\exp\left\{-\frac{n}{2}Q(\theta)\right\}d\theta}{ \int_{\Theta_{\delta,n}}\pi(\theta)d\theta}\leq Cn^{\frac{p_\gamma}{2}}\left( \exp\left\{-\frac{n}{2}\bar{Q}\right\}+\int_{C_2\|\gamma\|^2\geq \frac{c_n}{n}  }\exp\{-nC_1\|\gamma\|^2\}d\gamma\right)\leq\\
\leq o(1)+ \int_{C_2\|y\|^2\geq c_n }\exp\{-C_1\|y\|^2\}dy\to 0.
\end{align*}
In the last line we used the change of variables $y=\sqrt{n}\gamma$ and integrability of $\exp\{-\|y\|^2\}$. This proves (\ref{eq: posterior consistency}), and implies that for $\pi_{\Theta_{c_n}}$ the prior restricted to $\Theta_{c_n}$, the posterior $\pi_{\Theta_{c_n}}(\Upsilon|g_n)=\frac{\pi(\Upsilon\cap\Theta_{c_n}|g_n)}{\pi(\Theta_{c_n}|g_n)}$  defined on sets $\Upsilon\subseteq\Theta$ is asymptotically the same as $\pi(\Upsilon|g_n)$. If $\frac{c_n}{n}\to 0$, then due to compactness of $\Theta^*$, for large enough $n$ we have $\Theta_{c_n}\subseteq\Theta^*$. Thus, we can treat the parameterization described in Assumption \ref{ass:reparameterization} as applying to the whole parameter space $\Theta$.

The above implies that for any $\varepsilon>0$ there exists $\delta$ large enough such that
\begin{align}\label{eq: support of posterior}
P\left\{\sup_{\beta}n^{\frac{p_\gamma}{2}}\int_{\frac{\delta}{n}\leq\|\gamma\|^2} \exp\left\{-\frac{1}{2}Q_n(\beta,\gamma) \right\}\geq \varepsilon\right\}\leq \varepsilon,
\end{align}
and also $\sup_\beta P\left\{\|N(0,J^{-1}(\beta))\|\geq \delta\right\}<\varepsilon$.
Define $g_n^0(\beta,\gamma)=g_n(\beta)+\sqrt{n}\nabla(\beta)\gamma$ and $R_n(\beta,\gamma)=g_n(\beta,\gamma)-g_n^0(\beta,\gamma)$. Let us show that
\begin{align}\label{eq: VM theorem some statement}
\sup_\beta\sup_{\|\gamma\|^2\leq \frac{\delta}{n}}\|R_n(\beta,\gamma)\|\to^p 0.
\end{align}
Indeed, $\|R_n(\beta,\gamma)\|\leq \sqrt{n}\|\Phi(\beta,\gamma)-\nabla(\beta)\gamma\|+\|G_n(\beta,\gamma)-G_n(\beta)\|$. We have that $\sup_\beta\sup_{\|\gamma\|^2\leq \frac{\delta}{n}}\|G_n(\beta,\gamma)-G_n(\beta)\|\to^p0$ due to stochastic equicontinuity. Uniform differentiability implies $\sup_\beta\sup_{\|\gamma\|^2\leq \frac{\delta}{n}}\sqrt{n}\|\Phi(\beta,\gamma)-\nabla(\beta)\gamma\|\to 0.$

Denote  $Q_n^0(\beta,\gamma)=g^0_n(\beta,\gamma) \Sigma^{-1}(\beta)g^0_n(\beta,\gamma)$. Equation (\ref{eq: VM theorem some statement}) implies that
\begin{align}\label{eq: anothe VM statement}
\sup_\beta\sup_{\|\gamma\|^2\leq \frac{\delta}{n}}\left|1-\exp\left\{-\frac{1}{2}(Q_n(\beta,\gamma)-Q_n^0(\beta,\gamma))\right\}\right|\to^p0.
\end{align}
Indeed, the left-hand side is bounded above by
\begin{align*}
&\sup_\beta\sup_{\|\gamma\|^2
\leq \frac{\delta}{n}} |Q_n(\beta,\gamma)-Q_n^0(\beta,\gamma)|\leq\\ \leq &\sup_\beta\sup_{\|\gamma\|^2\leq \frac{\delta}{n}} \left\{ |(g_n+g_n^0)\widehat{\Sigma}_n^{-1}R_n|+|g_n^0(\widehat{\Sigma}_n^{-1}(\beta,\gamma)- \Sigma^{-1}(\beta))g_n^0|\right\}\to^p0.
\end{align*}
The last convergence follows from continuity of covariance function $\Sigma$, equation (\ref{eq: VM theorem some statement}), and boundedness in probability of $g_n$, $g_n^0$ and $\widehat\Sigma^{-1}$ over $\{\|\gamma\|^2\leq \frac{\delta}{n}\}$.

Denote $Q_n^\beta(\beta)=g_{n}\left(\beta\right)'M \left(\beta\right)g_{n}\left(\beta\right)$. Let us define a projection operator $P(\beta)=\Sigma^{-\frac{1}{2}}(\beta)\nabla(\beta)J(\beta)^{-1}\nabla(\beta)^\prime \Sigma^{-\frac{1}{2}}(\beta)$. Notice that $M(\beta)=\Sigma^{-\frac{1}{2}}(\beta)(I_k-P(\beta))\Sigma^{-\frac{1}{2}}(\beta)$.
\begin{align*}
Q_n^0(\beta,\gamma)=&g^0_n(\beta,\gamma)^\prime M(\beta)g^0_n(\beta,\gamma)+ g^0_n(\beta,\gamma)^\prime \Sigma^{-\frac{1}{2}}(\beta)P(\beta)\Sigma^{-\frac{1}{2}}(\beta)g^0_n(\beta,\gamma)=\\ = & Q_n^\beta(\beta)+\left(G^*(\beta)+\sqrt{n}\gamma\right)^\prime J\left(\beta\right)\left(G^*(\beta)+\sqrt{n}\gamma\right),
\end{align*}
where $G^*(\beta)=J(\beta)^{-1}\nabla(\beta)^\prime\Sigma^{-1}(\beta)G_n(\beta)$. Integration of the Gaussian pdf gives
\begin{align*}
&n^{\frac{p_\gamma}{2}}\int_{\|\gamma\|^2\leq\frac{\delta}{n}}\exp\left\{-\frac{1}{2}\left(G^*(\beta)+\sqrt{n}\gamma\right)^\prime J(\beta)\left(G^*(\beta)+\sqrt{n}\gamma\right)\right\}d\gamma=\\=&|J(\beta)|^{-\frac{1}{2}} P\left\{\left\|N\left(\frac{1}{\sqrt{n}}G^*(\beta),J^{-1}(\beta)\right)\right\|\leq \delta\right\}\to^p|J(\beta)|^{-\frac{1}{2}} P\left\{\|N\left(0,J^{-1}(\beta)\right)\|\leq \delta\right\},
\end{align*}
where $\delta$ was chosen large enough that $q(\delta)=P\left\{\|N(0,J^{-1}(\beta))\|\leq \delta\right\}\geq 1-\varepsilon$.
Thus,
\begin{align*}
\sup_{\beta}\left|n^{\frac{p_\gamma}{2}}\int_{\|\gamma\|^2\leq\frac{\delta}{n}}\exp\left\{-\frac{1}{2}Q^0_n(\beta,\gamma) \right\}d\gamma- |J(\beta)|^{-\frac{1}{2}}q(\delta)\exp\left\{-\frac{1}{2}Q_n^\beta(\beta)\right\}\right|\to^p 0.
\end{align*}
Joining together last statement with equations (\ref{eq: support of posterior}) and (\ref{eq: anothe VM statement}) we get
$$
\sup_{\beta}\left|n^{\frac{p_\gamma}{2}} \int_{\Gamma(\beta)}\exp\left\{-\frac{1}{2}Q_n(\beta,\gamma)\right\}d\gamma - |J(\beta)|^{-\frac{1}{2}}\exp\{-\frac{1}{2}Q_n^\beta(\beta)\}\right|\to^p0.
$$
Given statement (\ref{eq: support of posterior}), for $c(\beta,\gamma)$ satisfying assumptions of Theorem \ref{thm: consistency} we have
$$
\sup_{\beta}\left|n^{\frac{p_\gamma}{2}} \int_{\Gamma(\beta)}c(\beta,\gamma)\exp\left\{-\frac{1}{2}Q_n(\beta,\gamma)\right\}d\gamma - c(\beta,0)|J(\beta)|^{-\frac{1}{2}}\exp\left\{-\frac{1}{2}Q_n^\beta(\beta)\right\}\right|\to^p0.
$$
Assumption \ref{ass:reparameterization} implies  $\int_B\pi_\gamma(0|\beta)|J(\beta)|^{-\frac{1}{2}}\exp\{-\frac{1}{2}Q_n^\beta(\beta)\}d\pi(\beta)$ is stochastically bounded aways from zero. Thus,
 (\ref{eq: convergence of posterior integrals}) holds. $\Box$

\paragraph{Proof of Corollary \ref{cor: decisions}} For each $a\in \mathcal{A}$ we can apply (\ref{eq: convergence of posterior integrals}) to $c(\theta)=L(a,\theta)$. Since $L(a,\theta)$ is Lipschitz in $a$ and $\mathcal{A}$ is compact, this implies
$$
\sup_{a\in\mathcal{A}}\left|\int_\Theta L(a,\theta)d\pi(\theta|g_n)-\frac{\int_{B}L(a,\vartheta(\beta,0)) \exp\left\{-\frac{1}{2}Q_n^\beta(\beta)\right\}d\pi^0(\beta)}{\int_{B} \exp\left\{-\frac{1}{2}Q_n^\beta(\beta)\right\}d\pi^0(\beta)}\right|\to^p 0.
$$
We also have weak convergence of the process
\[
\frac{\int_{B}L(\cdot,\vartheta(\beta,0)) \exp\left\{-\frac{1}{2}Q_n^\beta(\beta)\right\}d\pi^0(\beta)}{\int_{B} \exp\left\{-\frac{1}{2}Q_n^\beta(\beta)\right\}d\pi^0(\beta)} \Rightarrow L(\cdot)
\]
on $\mathcal{A}$.
This implies $\int_\Theta L(\cdot,\theta)d\pi(\theta|g_n)\Rightarrow L(\cdot)$. Due to
 Theorem 3.2.2 in Van der Vaart and Wellner (1996),
$(s_n(g_n),s_n^0(g_n))\Rightarrow (\text{argmin}_{a\in\mathcal{A}}L(a),
\text{argmin}_{a\in\mathcal{A}}L(a)).
$
Thus, $s_n(g_n)-s_n^0(g_n)\to^p0$. $\Box$

\subsection*{Appendix B: Invariance Argument}

We seek default priors on $\mu$ that yield reasonable decision rules when combined with many different priors on $\theta^*$, including priors with restricted support $\widetilde\Theta\in\Theta_0$. For any such prior $\pi(\theta^*),$ define a corresponding restricted model  with parameter space $\widetilde\Theta\times\mathcal{H}_\mu$. Intuitively, by specifying a restricted-support prior $\pi(\theta^*)$, a researcher limits attention to the restricted model.

We next introduce a group of transformations.  Define $\mathcal{H}_{\mu,\widetilde\Theta}$ to be the linear subspace of functions in $\mathcal{H}_\mu$ that are zero everywhere on $\widetilde\Theta$,
$$
\mathcal{H}_{\mu,\widetilde\Theta}=\left\{\mu\in\mathcal{H}_\mu:\mu(\theta)=0\text{ for all }\theta\in\widetilde\Theta\right\}.
$$
The decision problem in the restricted model is invariant with respect to the
group of transformations that  for $\tilde\mu\in\mathcal{H}_{\mu,\widetilde\Theta}$ takes
$
\left(\xi,h\left(\cdot\right)\right)\to\left(\xi,h\left(\cdot\right)+\tilde{\mu}\left(\cdot\right)\right)
$
 in the sample space, and
$
\left(\theta^{*},\mu(\cdot)\right)\to\left(\theta^{*},\mu(\cdot)+\tilde{\mu}(\cdot)\right)
$
in the parameter space. See Chapter 3 of Lehmann and Casella (1998) for an introduction to invariance in decision problems.

A maximal invariant in the parameter space under this group of transformations is $\left(\theta^*,\mu\left(\cdot\right)\mathbb{I}\{\cdot\in\widetilde\Theta\}\right)$.
The statistic $\left(\xi,h\left(\cdot\right)\mathbb{I}\{\cdot\in\widetilde\Theta\}\right)$ is also invariant, and is sufficient for $\left(\theta^*,\mu\left(\cdot\right)\mathbb{I}\{\cdot\in\widetilde\Theta\}\right)$. Hence, once we restrict ourselves to invariant decision rules, it is without loss (in terms of attainable risk) to limit attention to decision rules that depend on the data only through $\left(\xi,h\left(\cdot\right)\mathbb{I}\{\cdot\in\widetilde\Theta\}\right)$.

By definition, Bayes decision rules in our setting minimize the quasi-posterior risk
$$
\min_{a\in\mathcal{A}}\frac{\int L(a,\theta)\ell^*(\theta)d\pi(\theta)}{\int\ell^*(\theta)d\pi(\theta)}
$$
 for almost every realization of the data.
 Motivated by the invariance of the restricted model, we seek default priors $\pi(\mu)$ such that for all priors $\pi(\theta^*)$ with restricted support $\widetilde\Theta\subset\Theta_0$, the joint prior $\pi(\theta^*)\pi(\mu)$ admits  Bayes decision rules depending only on $\left(\xi,h\left(\cdot\right)\mathbb{I}\{\cdot\in\widetilde\Theta\}\right)$.
\begin{definition}
A prior $\pi(\mu)$ is \textbf{invariance-compatible} for action space $\mathcal{A}$ and loss function $L$ if for all $\widetilde\Theta\subseteq \Theta_0$ and all priors $\pi(\theta^*)$ with support $\widetilde\Theta$, there exists a Bayes decision rule that depends on the data only through $\left(\xi,h\left(\cdot\right)\mathbb{I}\{\cdot\in\widetilde\Theta\}\right)$.
\end{definition}

Priors $\pi(\mu)$ such that $\ell^*(\theta^*)$ depends on the data only through $(\xi,h(\theta))$ are invariance-compatible for any $(\mathcal{A},L)$.  For unrestricted  action spaces and loss functions, the converse (up to scale) holds as well.

\begin{theorem}\label{prop: invariance}
A prior $\pi(\mu)$ is invariance-compatible for all $(\mathcal{A},L)$ pairs if and only if for any $\theta\in \Theta_0$ the integrated likelihood $\ell^{*}\left(\theta\right)$, defined in equation (6) in the main text,
depends on the data only through $\left(\xi,h\left(\theta\right)\right)$, up to scale.
\end{theorem}

For Gaussian process priors on $\mu,$ $\ell^{*}\left(\theta\right)$ corresponds to a Gaussian likelihood for $\xi$ with mean equal to the best linear predictor based on $h(\cdot)$.  In this case there is no scope for a data-dependent constant of proportionality, and $\pi(\mu)$ is invariance-compatible if and only if
$\ell^{*}\left(\theta\right)$ depends on the data only through $\left(\xi,h\left(\theta\right)\right)$.

\paragraph{Proof of Theorem \ref{prop: invariance}}

The ``if'' part of the statement is immediate.  For ``only if,'' first consider the case where $\widetilde\Theta=\left\{ \theta_{1},\theta_{2}\right\},$
$\mathcal{A}=\left\{ a_{1},a_{2}\right\} ,$ and
\[
L\left(a,\theta\right)=\begin{cases}
l_{j} & \text{if }\theta=\theta_{j},\text{ and }a\neq a_{j}\text{ for }j\in\left\{ 1,2\right\}; \\
0 & \text{otherwise},
\end{cases}
\]
so an incorrect decision incurs a loss $l_j$. Bayes decision rules must take
\[
a\in\begin{cases}
\{1\} & \text{if \ensuremath{\frac{\ell^{*}\left(\theta_{1}\right)}{\ell^{*}\left(\theta_{2}\right)} >\frac{l_{2}\pi\left(\theta_{2}\right)}{l_{1}\pi\left(\theta_{1}\right)}}};\\
\left\{ 1,2\right\}  & \text{if \ensuremath{\frac{\ell^{*}\left(\theta_{1}\right)}{\ell^{*}\left(\theta_{2}\right)}= \frac{l_{2}\pi\left(\theta_{2}\right)}{l_{1}\pi\left(\theta_{1}\right)}}};\\
\{2\} & \text{if \ensuremath{\frac{\ell^{*}\left(\theta_{1}\right)}{\ell^{*}\left(\theta_{2}\right)} <\frac{l_{2}\pi\left(\theta_{2}\right)}{l_{1}\pi\left(\theta_{1}\right)}}},
\end{cases}
\]
for almost every realization of the data.
Thus, we  see that the optimal action depends on the data only through
$\left(\xi,h\left(\theta_{1}\right),h\left(\theta_{2}\right)\right)$
for all values of $l_{1},$ $l_{2}$ if and only if the integrated
likelihood ratio $\ell^{*}\left(\theta_{1}\right)/\ell^{*}\left(\theta_{2}\right)$
depends on the data only through $\left(\xi,h\left(\theta_{1}\right),h\left(\theta_{2}\right)\right)$.

Since we can repeat this argument for all $\theta_{1},$ $\theta_{2}\in\Theta_0$,
we conclude that for $\pi(\mu)$ to be invariance-compatible, it must be the case that for all $\theta_{1},$ $\theta_{2}\in\Theta_0$
\[
\frac{\ell^{*}\left(\theta_{1}\right)}{\ell^{*}\left(\theta_{2}\right)}=\tilde{r} \left(\xi,h\left(\theta_{1}\right),h\left(\theta_{2}\right);\theta_{1},\theta_{2}\right),
\]
for some function $\tilde{r}$. This implies that for all $\theta_{1},$
$\theta_{2},$ $\theta_{3}\in\Theta_0$,
\[
\tilde{r}\left(\xi,h\left(\theta_{1}\right),h\left(\theta_{2}\right);\theta_{1},\theta_{2}\right)=
\frac{\tilde{r}\left(\xi,h\left(\theta_{1}\right),h\left(\theta_{3}\right);\theta_{1},\theta_{3}\right)}{\tilde{r}\left(\xi,h\left(\theta_{3}\right),h\left(\theta_{2}\right);\theta_{3},\theta_{2}\right)}.
\]
Hence, the right hand side does not depend on the value of $h\left(\theta_{3}\right)$, and in particular is the same as if $h\left(\theta_{3}\right)=0$. For a fixed value $\tilde\theta\in\Theta_0$, define
$
{r}\left(\xi,h\left(\theta_{1}\right),\theta_{1}\right)= \tilde{r}\left(\xi,h\left(\theta_{1}\right),0 ;\theta_{1},\tilde\theta\right),
$
and note that
\[
\ell^{*}\left(\theta_{1}\right)= \frac{\ell^{*}\left(\theta_{2}\right) }{{r}\left(\xi,h\left(\theta_{2}\right),\theta_{2}\right)} {r}\left(\xi,h\left(\theta_{1}\right),\theta_{1}\right),
\]
where the right side cannot depend on $\theta_2$. Thus
$
\ell^{*}\left(\theta\right)\propto {r}\left(\xi,h\left(\theta\right),\theta\right).
$
 $\Box$
 
\subsection*{Appendix C: Properties of Bayes and Quasi-Bayes Rules}

\paragraph{Admissibility of Proportional Prior Rules}
In finite-dimensional settings (specifically, settings where the covariance
function $\widetilde{\Sigma}$ has a finite number of nonzero eigenvalues)
choosing $\Omega=\lambda\widetilde{\Sigma}$ implies that $\pi\left(\mu\right)$
has support $\mathcal{H}_{\mu}$. In infinite-dimensional settings,
by contrast, $\pi\left(\mu\right)$ assigns probability zero to $\mathcal{H_{\mu}}$
-- see Section 3.1 of Berlinet and Thomas-Agnan (2004). It may not
therefore be obvious that the resulting Bayes decision rules are admissible.

This section shows that admissibility continues to hold under a weak
continuity condition. Specifically, for $\left\Vert \cdot\right\Vert $
the Euclidian norm, let $\left\Vert \mu\right\Vert _{\infty}=\sup_{\theta\in\Theta_{0}}\left\Vert \mu\left(\theta\right)\right\Vert $
be the sup norm, and define $\overline{\mathcal{H}}_{\mu}$ as the
closure of $\mathcal{H}_{\mu}$ under $\left\Vert \cdot\right\Vert _{\infty}.$
For a metric $d_{\theta}$ on $\Theta_{0}$, define 
\[
d\left(\left(\theta,\mu\right),\left(\theta',\mu'\right)\right)=d_{\theta}\left(\theta,\theta'\right)+\left\Vert \mu-\mu'\right\Vert _{\infty}
\]
on $\Theta_{0}\times\overline{\mathcal{H}}_{\mu}.$ Let $\mathcal{S}$
be the class of decision rules whose risk functions $\mathbb{E}_{\theta,\mu}\left[L\left(s\left(g\right),\theta\right)\right]$
are continuous with respect to $d$.

\begin{theorem}\label{thm: admissibility}For $\pi\left(\theta,\mu\right)=\pi\left(\theta\right)\pi\left(\mu\right)$,
where $\pi\left(\theta\right)$ has support $\Theta_{0}$ and $\pi\left(\mu\right)$
corresponds to $\mu\sim\mathcal{GP}\left(0,\lambda\widetilde{\Sigma}\right)$
with $\lambda>0$, any decision rule $s_{\pi}\in\mathcal{S}$ with
\begin{equation}
\int \mathbb{E}_{\theta,\mu}\left[L\left(s_{\pi}\left(g\right),\theta\right)\right]d\pi\left(\theta,\mu\right)=\min_{s\in\mathcal{S}}\int \mathbb{E}_{\theta,\mu}\left[L\left(s\left(g\right),\theta\right)\right]d\pi\left(\theta,\mu\right)\label{eq: Bayes optimality}
\end{equation}
 is admissible relative to $\mathcal{S}$ and parameter space $\Theta_{0}\times\mathcal{H}_{\mu}$.
\end{theorem}

\paragraph{Proof of Theorem \ref{thm: admissibility}}
Suppose there exists $\tilde{s}\in\mathcal{S}$
that dominates $s_{\pi}.$ Since $\tilde{s}$ dominates $s_{\pi},$
and the risk function is continuous, the set
\begin{equation}
\left\{ \left(\theta,\mu\right)\in\Theta\times\overline{\mathcal{H}}_{\mu}:\mathbb{E}_{\theta,\mu}\left[L\left(s_\pi\left(g\right),\theta\right)\right]-\mathbb{E}_{\theta,\mu}\left[L\left(\tilde{s}\left(g\right),\theta\right)\right]>0\right\} \label{eq: preimage}
\end{equation}
is nonempty and open, while the set
\begin{equation}
\left\{ \left(\theta,\mu\right)\in\Theta\times\overline{\mathcal{H}}_{\mu}:\mathbb{E}_{\theta,\mu}\left[L\left(s_\pi\left(g\right),\theta\right)\right]-\mathbb{E}_{\theta,\mu}\left[L\left(\tilde{s}\left(g\right),\theta\right)\right]<0\right\} \label{eq: negative preimage}
\end{equation}
 is empty. By Lemma 5.1 in van der Vaart and van Zanten (2008), $\pi\left(\mu\right)$
has support $\overline{\mathcal{H}}_{\mu}$. From the definition of
$d$, $\pi\left(\theta,\mu\right)$ must therefore assign positive
mass to (\ref{eq: preimage}). Since (\ref{eq: negative preimage})
is empty, this implies that $\int \mathbb{E}_{\theta,\mu}\left[L\left(s_{\pi}\left(g\right),\theta\right)\right]d\pi\left(\theta,\mu\right)$
strictly exceeds $\int \mathbb{E}_{\theta,\mu}\left[L\left(\tilde{s}\left(g\right),\theta\right)\right]d\pi\left(\theta,\mu\right)$.
We have obtained a contradiction with (\ref{eq: Bayes optimality}).

\paragraph{Limit-of-Bayes}
We obtain the quasi-posterior of Chernozhukov
and Hong (2003) as the limit of a sequence of posteriors for proper
priors. Here, we show that under the conditions similar to Theorem
\ref{Thm: Brown admissibility}, quasi-Bayes decision rules are likewise
the pointwise limit of the corresponding Bayes decision rules, and
that the same holds for their risk functions. To state this result,
consider any sequence of finite values $\lambda_{r}\to\infty$ as $r\to\infty$, define
$\pi_{r}\left(\theta,\mu\right)=\pi\left(\theta\right)\pi_{r}\left(\mu\right)$
to be the corresponding sequence of priors, and let $s_{\pi_{r}}$
be the corresponding sequence of Bayes decision rules,
\[
s_{\pi_{r}}\left(g\right)\in\text{argmin}_{a\in\mathcal{A}}\frac{\int L\left(a,\theta\right)\ell_r^{*}\left(\theta\right)d\pi\left(\theta\right)}{\int\ell_r^{*} \left(\theta\right)d\pi\left(\theta\right)}=\text{argmin}_{a\in\mathcal{A}}\int L\left(a,\theta\right)\ell_r^{*}\left(\theta\right)d\pi\left(\theta\right),
\]
and $s_{\pi_{\infty}}\left(g\right)$ the quasi-Bayes decision rule.

\begin{theorem}\label{thm: limit-of-Bayes} Suppose that $\mathcal{A}$
is compact and convex, that $L\left(a,\theta\right)$ is uniformly
bounded as well as continuous and strictly convex in $a$ for every
$\theta$, and that $\Sigma\left(\theta,\theta\right)$ is everywhere
full rank. Then $s_{\pi_{r}}\left(g\right)\to s_{\pi_{\infty}}\left(g\right)$
for every $g$ and
\begin{equation}
\mathbb{E}_{\theta,\mu}\left[L\left(s_{\pi_{r}}\left(g\right),\theta\right)\right]\to \mathbb{E}_{\theta,\mu}\left[L\left(s_{\pi_{\infty}}\left(g\right),\theta\right)\right]\text{ for each \ensuremath{\left(\theta,\mu\right)\in\Theta_{0}\times\mathcal{H}_{\mu}}}.\label{eq: Risk convergence}
\end{equation}
\end{theorem}

\paragraph{Proof of Theorem \ref{thm: limit-of-Bayes}}

Recall that $\ell_r^{*}\left(\theta\right)=\left|\Lambda_{r}(\theta)\right|^{-\frac{1}{2}}\cdot\exp\left(-\frac{1}{2}u_{r}(\theta)'\Lambda_{r}(\theta)^{-1}u_{r}(\theta)\right),$
where $u_{r}(\theta)=\frac{\lambda_{r}}{1+\lambda_{r}}\psi\left(\theta\right)^{-1}g\left(\theta\right)+\frac{1}{1+\lambda_{r}}\xi$,
$\Lambda_{r}(\theta)=\frac{\lambda_{r}}{1+\lambda_{r}}\left[\psi\left(\theta\right)^{-1}\right]\Sigma\left(\theta,\theta\right)\left[\psi\left(\theta\right)^{-1}\right]'+\frac{1}{1+\lambda_{r}}Var(\xi).$
Since the minimum eigenvalue of $\Sigma\left(\theta,\theta\right)$
is bounded away from zero by compactness of $\Theta_0$, and $g\left(\cdot\right)$ is bounded, we have:
\[
\ell_{r}^{*}\left(\theta\right)\to\ell_{\infty}^{*}\left(\theta\right)=\left|\psi(\theta)\right|\cdot\left|\Sigma(\theta,\theta)\right|^{-\frac{1}{2}}\cdot\exp\left(-\frac{1}{2}g(\theta)'\Sigma(\theta,\theta)^{-1}g(\theta)\right)
\]
uniformly on $\Theta_{0}$. Since the loss function is bounded,
\[
\int L\left(a,\theta\right)\ell_{r}^{*}\left(\theta\right)d\pi\left(\theta\right)\to\int L\left(a,\theta\right)\ell_{\infty}^{*}\left(\theta\right)d\pi\left(\theta\right)\text{ uniformly on }\mathcal{A},
\]
which, since the loss is strictly convex, implies that $s_{\pi_{r}}\left(g\right)\to s_{\pi_{\infty}}\left(g\right).$
Finally, (\ref{eq: Risk convergence}) follows from another application
of the dominated convergence theorem.
\subsection*{Appendix D: Simulation Design and Additional Results. }

\paragraph{Simulation Design}
This section describes the simulation design used to produce Figure \ref{fig: q75 approximation} in Section \ref{sec: weak ID}. We base our simulations on the Graddy (1995) data. In particular,
for our data-calibrated simulations, we:
\begin{enumerate}
\item Estimate $\hat{\theta}$ using continuously updating GMM.
\item Draw $\left(W,Y,Z\right)$ from the empirical distribution, and generate
new outcomes $Y^{*}$ by adding normal noise with standard deviation
equal to one-tenth that of $Y,$ i.e. $Y^{*}=Y+\varepsilon$, with
$\varepsilon\sim N\left(0,\frac{1}{100}\widehat{Var}\left(Y\right)\right).$
\item Use exponential tilting for find weights $\omega_{i}$ on the observations
$\left\{ W_{i},Y_{i},Z_{i}\right\} $ in the original sample such
that the quantile IV moments, evaluated with the observations constructed
in step 2, hold exactly at $\hat{\theta}$.
\item Draw samples $\left\{ W,Y^{*},Z\right\} $ with sampling weights $\omega_{i}$ as in step 3,
and outcomes $Y^{*}$ as in step 2.
\end{enumerate}
This construction ensures that (i) $Y$ is continuously distributed,
as necessary for conditional quantiles to be uniquely defined and
(ii) the (over-identified) GMM moments hold exactly at the value
$\hat{\theta}$ estimated on the original data. Denote the resulting
distribution by $P^{*}$.

To construct the distribution $P_{0},$ we take the distribution $P^{*}$
and, as described in Section \ref{sec: limit experiment}, multiply all entries
of $Z$ but the constant by a mean-zero (specifically, Rademacher)
random variable. This construction ensures that $P_{0}$ dominates
$P^{*}$ and, as discussed in the text, that $\theta^{*}$ is set-identified
under $P_{0}$.
To construct $P_{n,f}$, we draw from a mixture between $P_{0}$
and $P^{*}$, with weight $\sqrt{\frac{111}{n}}$ on $P^{*}$. Hence,
for the original sample size ($n=111$) $P_{n,f}=P^{*}$, while as
$n\to\infty$, $P_{n,f}$ converges to $P_{0}$. In particular, $P_{n,f}$
satisfies equation (\ref{eq: Path restriction}) for $f\propto\frac{dP^*}{dP_0}-1.$
Finally, to compute GMM estimates in each draw of simulated data, we run an MCMC
algorithm (specifically slice sampling), and report the parameter values yielding the minimal objective function value encountered as our estimate.

\begin{figure}
\includegraphics[scale=0.45]{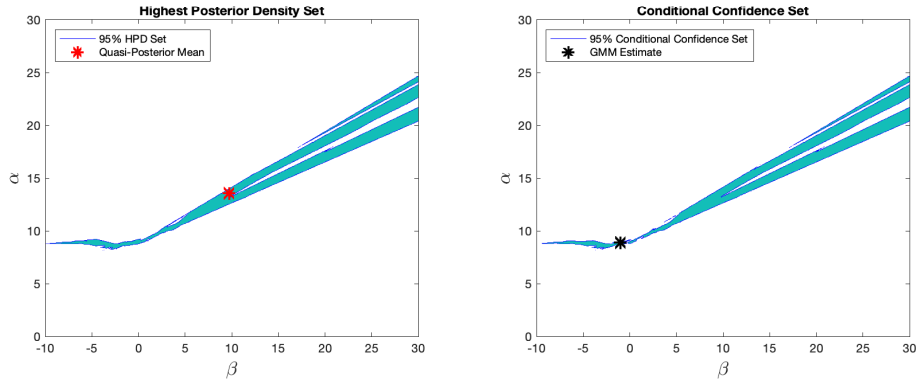}
\caption{Quasi-Bayes 95\% highest posterior density set, and 95\% conditional frequentist confidence set, for $\tau=0.75$ based on Graddy (1995) data\label{fig: CS plots}}
\end{figure}

\paragraph{Additional Empirical Results}
Figure \ref{fig: CS plots} plots the 95\% conditional confidence set formed by inverting our weighted average power optimal tests, along with the 95\% highest posterior density set. The conditional confidence sets do have asymptotically correct frequentist coverage and use quasi-posterior only to form a powerful test statistics. The 95\% highest posterior density set does not have frequentist guarantees and is a Bayesian object. In this application the two sets have a quite similar shape, but the  confidence
set is slightly smaller, covering 4.74\% of the parameter space as compared to 4.82\% for the highest posterior density set.

\begin{figure}
\includegraphics[scale=0.4]{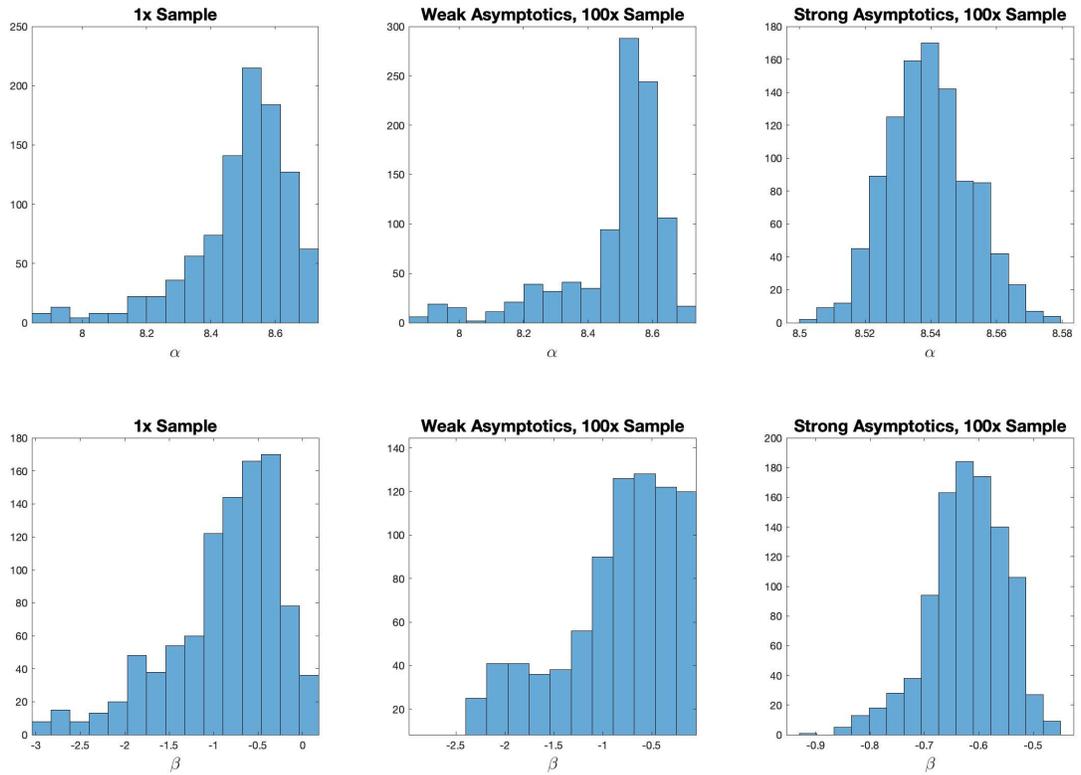}
\caption{Finite sample distribution of GMM estimator in simulations calibrated to Graddy (1995) data, along with weak- and strong-asymptotic large-sample distributions.  Based on 1000 simulation draws.\label{fig: q50 approximation}}
\end{figure}

Figure \ref{fig: q50 approximation} plots the small sample distribution of the GMM estimate for the median, $\tau=0.5$ and weak- and large-sample asymptotic distributions.  We again see that the weak-asymptotic approximation appears substantially more accurate.

  \end{document}